\documentclass{aa}
\usepackage{graphics,epsfig,txfonts}
\usepackage{enumerate}
\usepackage{rotating}
\usepackage{caption} 
\usepackage{subfigure}

\usepackage{lscape}
\usepackage{longtable} 
\usepackage{natbib}
\bibpunct{(}{)}{;}{a}{}{,}

\newcommand{\ergcm}[1]{$\times 10^{#1}$ erg cm$^{-2}$ s$^{-1}$}
\newcommand{\oergcm}[1]{$10^{#1}$ erg cm$^{-2}$ s$^{-1}$}
\newcommand{\ergs}[1]{$\times 10^{#1}$ erg s$^{-1}$}

\newcommand{\nh}{$\ensuremath{N_{\mathrm{H}}}$}

\newcommand{\Hone}{\ion{H}{I}}

\newcommand{\ltsima}{$\buildrel < \over \sim$}
\newcommand{\lsim}{\lower.5ex\hbox{\ltsima}}
\newcommand{\gtsima}{$\buildrel > \over \sim$}
\newcommand{\gsim}{\lower.5ex\hbox{\gtsima}}
\newcommand{\xmm}{{\it XMM-Newton}}

\def\rahour{\hbox{\ensuremath{^{\rm h}}}}
\def\ramin{\hbox{\ensuremath{^{\rm m}}}}
\def\rasec{\hbox{\ensuremath{^{\rm s}}}}

\usepackage{color}

\usepackage[switch]{lineno}

\begin{document}
 
\title{Identification of AGN in the \xmm\ X-ray survey of the SMC
       \thanks{Based on observations with 
               XMM-Newton, an ESA Science Mission with instruments and contributions 
               directly funded by ESA Member states and the USA (NASA)}
       \thanks{Based on observations made with VISTA at the Paranal Observatory under programme ID(s) 179.B-2003(G), 179.B-2003(F), 179.B-2003(D), 179.B-2003(C), 179.B-2003(H), 179.B-2003(B)}
      }

\author{C.~Maitra\inst{1}\thanks{\email{cmaitra@mpe.mpg.de}} \and
        F.~Haberl\inst{1} \and
        V.~D.~Ivanov\inst{2,3}
        Maria-Rosa L. Cioni\inst{4} \and
        Jacco Th. van Loon\inst{5}
       }

\titlerunning{AGN behind the SMC}
\authorrunning{Maitra et al.}

\institute{Max-Planck-Institut f\"{u}r extraterrestrische Physik, Giessenbachstra{\ss}e, 85748 Garching, Germany
        \and European Southern Observatory, Ave. Alonso de Córdova 3107, Vitacura, Santiago, Chile
	\and European Southern Observatory, Karl-Schwarzschild-Str. 2, 85748 Garching bei M\"{u}nchen, Germany
        \and Leibniz-Institut f\"{u}r Astrophysik Potsdam, An der Sternwarte 16, 14482 Potsdam, Germany
        \and Lennard-Jones Laboratories, Keele University, ST5 5BG, UK 
	   }


\abstract
{Finding Active Galactic Nuclei (AGN) behind the Magellanic Clouds (MCs) is difficult because of the high stellar density in these fields. 
Although the first AGN behind the Small Magellanic Cloud (SMC) were reported in the 1980s, it is only recently that the number of AGN 
known behind the SMC has increased by several orders of magnitude.} 
{The mid-infrared colour selection technique has been proven to be an efficient means of identifying AGN, especially obscured sources. 
The X-ray regime is complementary in this regard and we use \xmm\ observations to support the identification of AGN behind the SMC.}
{We present a catalogue of AGN behind the SMC by correlating an updated X-ray point source catalogue from our \xmm\ survey of the SMC 
with already known AGN from the literature as well as a list of candidates obtained from the ALLWISE mid-infrared colour selection criterion. 
We studied the properties of the sample with respect to their redshifts, luminosities and X-ray spectral 
characteristics. We also identified the near-infrared counterpart of the sources from the VISTA observations. 
}
{The redshift and luminosity distributions of the sample (where known) indicate that we detect sources from nearby Seyfert galaxies to distant and obscured quasars. The X-ray hardness ratios are compatible with those typically expected for AGN. The VISTA colours and variability are also consistent in 
this regard. A positive correlation was observed between the integrated X-ray flux (0.2--12 keV) and the ALLWISE and VISTA magnitudes. 
We further present a sample of new candidate AGN and candidates for obscured AGN. 
All of these make an interesting subset for further follow-up studies.
An initial spectroscopic follow-up of 6 out of the 81 new candidates showed all six sources are active galaxies, albeit two with narrow emission lines.}
{}

 
\maketitle

\section{Introduction}
\label{sec:introduction}

Active galactic nuclei (AGN) are amongst the most powerful and steady sources of luminosity in the universe ranging from AGN in nearby galaxies
emitting at luminosities
about 10$^{40}$ erg s$^{-1}$, to distant quasars emitting $>$ 10$^{47}$ erg s$^{-1}$. AGN play a crucial role in the formation and evolution
of galaxies \citep[e.g.][]{2014IJMPD..2330015C}, and in
the growth of supermassive black holes \citep{1999MNRAS.303L..34F}. 

Identification of AGN behind the Magellanic Clouds (MCs) can be a challenging task, because of the high stellar density in these fields.
The first small samples of quasars behind the Small Magellanic Cloud (SMC) were reported in the 1980s \citep{1982MNRAS.200.1007M,1983PASAu...5....2W}. 
\citet{1997MNRAS.285..111T} confirmed eight new sources by performing optical spectroscopic follow-up of {\it ROSAT} X-ray sources. Further, 
\citet{2003AJ....126..734D,2003AJ....125.1330D} and \citet{2003AJ....125....1G} identified five candidates from X-ray selection and five using the 
optical variability properties from the Magellanic Cloud OGLE-II data \citep{2002AcA....52..241E}. 
The Magellanic Quasar Survey  (MQS) played a significant role in increasing the number of AGN behind the MCs. 
In the first work of a series, \citet{2009ApJ...701..508K} selected 657 candidates using {\it Spitzer} space telescope in the infrared and near-infrared photometry. 
Subsequently, based on the candidates from \citet{2009ApJ...701..508K}, and additional ones selected by optical variability from OGLE data
together with mid-IR and/or X-ray properties,
\citet[][OGLE-II]{2011ApJS..194...22K} and \citet[][OGLE-III]{2013ApJ...775...92K} were able to confirm 193 of the 766 candidates selected in total by performing follow-up spectroscopy.

The last years have witnessed an increase in the number of identified AGN by several orders of magnitude. 
A large fraction of this owes to the advent of the mid-infrared colour selection technique which has proven to be a very efficient means of identifying AGN.
This technique is based on detecting the hot obscured dust surrounding an AGN, which is much less affected by extinction. 
For the same reason this method is also effective in finding the most obscured AGN, and is in a way complementary to X-ray surveys of AGN at 
energies $<$10 keV which are only partially sensitive to Compton-thick AGN \citep[\nh $>10^{24}$ cm$^{-2}$;][]{2008A&A...487..119D}. 
The mid-infrared selection techniques have been introduced by \citet{2004ApJS..154..166L}, \citet{2005ApJ...631..163S} and \citet{2012ApJ...748..142D} 
using data from the {\it Spitzer} mission. This was later adapted  for the
Wide-field Infrared Survey Explorer \citep[{\it WISE};][]{2010AJ....140.1868W} mission which performed an all-sky survey in the infrared bands at 3.4, 4.6, 12 and 22 $\mu$m.
The mission achieved $5\sigma$ point source sensitivities better than 0.08, 0.11, 1 and 6 mJy at 3.4, 4.6, 12 and 22 $\mu$m respectively, with angular resolutions of
6.1, 6.4, 6.5 and $12\rlap{.}\arcsec0$ in the respective bands.
\citet{2012ApJ...753...30S,2013ApJ...772...26A} applied one colour selection criterion using 3.4 and 4.6 $\mu$m wavebands from the ALLWISE data to select AGN. 
\citet{2012MNRAS.426.3271M} formulated a more advanced and effective two-colour wedge selection criterion of these sources.
All of the selection criteria are based on the fact that the AGN separate cleanly from stars and star forming galaxies in the mid-infrared colour space.

Based on this, \citet{2015ApJS..221...12S} presented an all-sky catalogue of 1.4 million AGN selected using
the two-colour mid-infrared criteria for AGN \citep{2012MNRAS.426.3271M}, applied to sources from the {\it WISE} final catalogue release (ALLWISE).
This included approximately 1.1 million previously uncatalogued AGN.
Additionally, the Half-Million Quasars catalogue (HMQ) \citep{2015PASA...32...10F} lists a sample of 510,764 objects. This included high confidence SDSS based photometric quasars 
with radio/X-ray associations as well as BL Lac and type 2 objects with accurate source positions. 
Recently, \citet{2017yCat.7277....0F} published the Million Quasars (MILLIQUAS) catalogue, which is an updated version of the former catalogue including $\sim$900k 
high-confidence quasar candidates from SDSS-based photometric quasar catalogues and also from all-sky radio/X-ray associated objects, bringing the total count to 
1,422,219. However only a small fraction of this catalogue contained X-ray selected sources or sources with X-ray associations.

The \xmm\ survey of the SMC \citep{2012A&A...545A.128H} provided the deepest complete coverage of the main body of the SMC (bar and eastern wing, $\sim5.58$ deg$^2$) in the 
X-ray (0.2$-$12.0 keV) band. 
The bulk ($>$71\%) of the 3053 X-ray-emitting point sources that were detected are expected to be AGN although only 72 could be identified with high confidence 
\citep[][]{2013A&A...558A...3S}.
This is because, for an individual faint ($\sim$\oergcm{-14}) X-ray source, it is difficult to prove the AGN nature. Moreover, owing to the coordinate 
uncertainties of an \xmm\ source (typical $\sigma\sim1\rlap{.}\arcsec2$, after astrometric correction), it is challenging to find the optical counterpart because of the high density of SMC stars, which causes 
chance coincidences with typically up to $\sim$4 optical sources. \cite{2013A&A...558A.101S} used associations of the X-ray point sources with radio emission to effectively 
select 43 new candidates for AGN behind the SMC. With the large database of all-sky quasar catalogues available now, and additional effective selection methods in other wavebands 
(especially in the mid-infrared) well tested, we have now the ideal opportunity to identify many more quasars behind the SMC. 

AGN behind the MCs are particularly important for two additional reasons. 
First, they are ideally suited as anchors for an absolute astrometric reference system for proper motion studies \citep[e.g. ][]{2013ApJ...764..161K,2014A&A...562A..32C}.
Also in the context of X-ray sources, we note the importance of having identified X-ray sources to improve the astrometric quality of X-ray catalogues in the field of the MCs.
The brightest sources might be used to probe absorption by the interstellar medium in the MCs.
Finally, the AGN can be the subject of investigations using the comprehensive multi-epoch and multi-wavelength data available for the MCs \citep[e.g.][]{2009ApJ...698..895K}.

In this paper we present a catalogue of X-ray detected AGN behind the SMC by correlating an updated point source catalogue of the \xmm\ SMC survey 
with lists of AGN known from literature and of candidates obtained in this work using the ALLWISE mid-infrared colour selection criterion. 
In Sect. 2 we describe the observations, the data reduction and the production of the AGN samples based on mid-infrared selections and from published catalogues and their correlation with the X-ray data. In Sect. 3 we describe the results and characterise the selected AGN. In Sect. 4 we present a summary and conclusions. For calculation of luminosities,  we adopted the standard cosmological parameters:
$\Omega_{M}=0.3$, $\Omega_{A}=0.7$ and $H_{0}$ =70 km s$^{-1}$ Mpc$^{-1}$.

%

%

\section{Observations and data reduction}
\label{sec:observations}

\subsection{X-ray data and the updated SMC survey}
The \xmm\ \citep{2001A&A...365L...1J} survey of the SMC \citep{2012A&A...545A.128H} provided the deepest complete coverage of the main body of the SMC 
(bar and eastern wing, $\sim$5.58 deg$^2$) in the 0.2$-$12.0 keV band.
Using additional outer fields, a catalogue of 3053 X-ray point sources was created. 
A detailed description of the catalogue and the used observations are presented in \citet{2013A&A...558A...3S}.
We added to this sample new observations of the SMC wing region acquired as part of the SMC Wing survey (PI: F. Haberl), and all observations publicly available up 
to satellite revolution 3177 (2017-04-14).
This resulted in 44 new pointings (including archival observations). 
The updated survey has a total exposure of 3.4 Ms and covers an area of $\sim$6.67 deg$^2$ and will be described in detail in a forthcoming paper.
The new observations are listed in Table~\ref{tab:xmm_obs}.
We reprocessed the entire dataset mentioned above in the same way as for the SMC X-ray point-source catalogue using the latest version of SAS (16.1), 
also including new boresight corrections utilising the catalogue of AGN compiled from this work. 
In contrast to \citet{2013A&A...558A...3S} we applied background filtering individually to each instrument. This resulted in a gain of exposure times for
MOS1 and MOS2 in most cases.
For the compilation of the updated SMC point source catalogue we applied a reduced systematic error of $0\rlap{.}\arcsec33$ in comparison to \citet{2013A&A...558A...3S}
who used a value of $0\rlap{.}\arcsec5$. 
This is comparable to the systematic error applied to the 3XMM-DR6 catalogue \citep{2016A&A...590A...1R}\footnote{After resolving a known problem in coordinate 
conversion from SAS 15.0 onwards, a better agreement between the X-ray and optical positions were obtained. 
See https://www.cosmos.esa.int/web/xmm-newton/sas-release-notes-1500}.
A total of 8690 detections were found from the entire SMC survey area, which include detections with detection likelihood of $\geq$ 6. 
Unique sources were included in the catalogue when at least one detection was found with a detection likelihood of $\geq$ 10. 
This resulted in 4449 unique X-ray point sources. 
The master source positions and the corresponding errors were calculated from the error-weighted average of the individual detections. 
All the sources identified in this work were screened individually by visual inspection to avoid spurious detections.  

\begin{table*}
\caption[]{New \xmm\ observations since \citet{2013A&A...558A...3S} given in chronological order.}
\label{tab:xmm_obs}
\centering
\begin{tabular}{lrrrrrrr}
\hline\hline\noalign{\smallskip}
\multicolumn{1}{c}{ObsID} &
\multicolumn{1}{c}{R.A.} &
\multicolumn{1}{c}{Dec.} &
\multicolumn{1}{c}{Exp$_{\rm pn}$} &
\multicolumn{1}{c}{Exp$_{\rm m1}$} &
\multicolumn{1}{c}{Exp$_{\rm m2}$} &
\multicolumn{1}{c}{$\Delta$R.A.} &
\multicolumn{1}{c}{$\Delta$Dec.} \\
\noalign{\smallskip}\hline\noalign{\smallskip}
0412981301 & 01:04:22.48 & $-$72:01:10.8 & 2873 & 21683 & 22730 &	1.197 & 0.659 \\
0677980301 & 01:01:37.89 & $-$72:25:28.4 & 5233 & 9433 & 11050 &	$-$0.479 & 0.019 \\
0412981401 & 01:03:42.98 & $-$72:01:29.2 & 28721 & 32744 & 33634 &	$-$2.062 & $-$0.247 \\
0679180301 & 00:23:52.40 & $-$72:23:09.3 & 9538 & 14103 & 14576 &	-- & -- \\
0412981501 & 01:04:26.97 & $-$72:00:52.6 & 0 & 29924 & 29929 &	$-$0.878 & $-$0.115 \\
0412981601 & 01:02:07.05 & $-$71:55:32.2 & 0 & 0 & 28715 &	0.384 & $-$0.151 \\
0700381801 & 01:28:05.26 & $-$73:31:59.6 & 29880 & 31458 & 31463 &	0.106 & 0.362 \\
0693050501 & 01:23:20.21 & $-$75:21:21.2 & 9449 & 13621 & 14128 &	1.670 & 0.632 \\
0412981701 & 01:04:09.40 & $-$72:01:02.2 & 47095 & 56783 & 43900 &	$-$0.375 & $-$1.010 \\
0721960101 & 01:28:03.92 & $-$73:31:51.2 & 70282 & 77693 & 78552 &	1.521 & 0.264 \\
0724650301 & 01:33:15.19 & $-$74:25:01.8 & 22628 & 33114 & 33621 &	0.857 & 1.387 \\
0700580101 & 00:57:06.59 & $-$72:24:37.3 & 8847 & 13449 & 13419 &	0.151 & $-$1.199 \\
0412982101 & 01:04:08.95 & $-$72:00:44.6 & 30371 & 31953 & 31923 &	0.039 & $-$1.102 \\
0700580401 & 00:57:22.99 & $-$72:26:09.6 & 14005 & 19277 & 19451 &	0.547 & $-$0.777 \\
0700580601 & 00:57:18.04 & $-$72:25:24.1 & 16377 & 18646 & 18619 &	$-$0.002 & $-$0.452 \\
0674730201 & 00:59:00.25 & $-$71:40:09.0 & 19199 & 19395 & 19303 &	-- & -- \\
0412982301 & 01:04:24.13 & $-$72:01:40.5 & 43350 & 43533 & 43508 &	2.898 & $-$0.837 \\
0412982201 & 01:04:09.51 & $-$72:00:44.4 & 32418 & 33596 & 33624 &	0.970 & $-$0.209 \\
0741450101 & 01:25:40.47 & $-$73:17:59.2 & 49673 & 50199 & 76693 &	$-$0.433 & $-$1.783 \\
0770580701 & 00:54:50.45 & $-$73:39:49.4 & 0 & 9650 & 9355 &	2.211 & $-$1.208 \\
0770580801 & 00:54:51.89 & $-$73:39:58.8 & 28080 & 28395 & 28397 &	1.012 & $-$0.183 \\
0764780201 & 00:43:02.49 & $-$73:39:45.3 & 46365 & 47950 & 47921 &	0.311 & $-$0.977 \\
0770580901 & 00:54:54.88 & $-$73:40:22.9 & 20395 & 22027 & 23663 &	0.600 & $-$3.385 \\
0764050101 & 00:54:19.22 & $-$72:29:45.3 & 13141 & 26649 & 26619 &	0.057 & 0.585 \\
0412982501 & 01:04:23.44 & $-$72:01:36.4 & 33362 & 33535 & 33524 &	$-$0.104 & $-$0.053 \\
0412982401 & 01:04:09.37 & $-$72:00:45.4 & 32993 & 35803 & 36770 &	0.743 & $-$0.315 \\
0763590401 & 01:22:03.92 & $-$72:57:08.2 & 21034 & 23543 & 23515 &	0.022 & 1.810 \\
0764050201 & 00:53:40.14 & $-$72:31:28.4 & 17667 & 30099 & 33086 &	$-$0.276 & $-$0.654 \\
0791580701 & 01:03:40.16 & $-$72:02:17.0 & 30475 & 30651 & 30611 &	$-$0.513 & 0.615 \\
0791580801 & 01:03:40.12 & $-$72:02:18.0 & 12128 & 12655 & 12615 &	$-$0.498 & 0.240 \\
0791580901 & 01:03:40.03 & $-$72:02:17.4 & 11777 & 12603 & 12615 &	$-$0.803 & 0.579 \\
0791581001 & 01:03:40.12 & $-$72:02:18.2 & 30477 & 30655 & 30615 &	$-$0.454 & $-$0.287 \\
0791581101 & 01:03:40.13 & $-$72:02:18.1 & 13815 & 14292 & 14258 &	$-$0.560 & $-$0.511 \\
0791581201 & 01:03:40.04 & $-$72:02:17.1 & 350 & 14973 & 15659 &	$-$1.040 & 0.113 \\
0784690601 & 01:29:13.99 & $-$73:07:54.7 & 30337 & 34546 & 34620 &	$-$1.723 & $-$1.465 \\
0784690701 & 01:23:14.81 & $-$73:29:11.7 & 27660 & 32846 & 33187 &	1.822 & 0.537 \\
0784690401 & 01:19:18.91 & $-$73:39:47.7 & 24747 & 37517 & 37664 &	1.380 & $-$0.545 \\
0412983201 & 01:04:09.26 & $-$72:00:46.7 & 29216 & 33323 & 33400 &	0.989 & $-$1.236 \\
0784690801 & 01:33:12.21 & $-$73:17:02.4 & 38141 & 41388 & 41619 &	1.142 & $-$0.077 \\
0784690201 & 01:12:38.71 & $-$73:28:14.2 & 29605 & 36502 & 37609 &	$-$0.135 & $-$1.117 \\
0412983301 & 01:04:23.86 & $-$72:01:49.2 & 29036 & 33585 & 33617 &	0.996 & 1.362 \\
0784690101 & 01:09:32.19 & $-$73:19:19.2 & 22309 & 27317 & 28281 &	2.955 & 1.665 \\
0784690301 & 01:15:02.69 & $-$73:43:27.9 & 38684 & 40282 & 40259 &	$-$0.144 & $-$1.854 \\
0803210201 & 01:07:23.59 & $-$72:28:27.2 & 33739 & 35649 & 35621 &	$-$0.588 & 0.326 \\

 \noalign{\smallskip}\hline
\end{tabular}
\tablefoot{Coordinates are in J2000. Net exposures for all EPIC instruments are given in s.
$\Delta$R.A. and $\Delta$Dec denote the shifts in R.A. and Dec. after boresight correction in arcsec. 
For the observation log of the \xmm\ SMC catalogue, see Table B.1 of \citet{2013A&A...558A...3S}.
Observations starting with 04129 correspond to routine calibration observations of 1ES0102--72 and 07846
to the SMC wing survey.}
\end{table*}

\subsection{The mid infrared sample \& the ALLWISE data}
We selected all sources from the ALLWISE catalogue within a radius of 3.33\degr\ around R.A. = 01\rahour00\ramin00\rasec,
Dec. = $-$72\degr30\arcmin00\arcsec\ (area of the \xmm\ SMC survey)
following the two-colour selection criterion of \citet[][see eqn. 3 and 4]{2012MNRAS.426.3271M}. We additionally imposed a
requirement that 
the selected sources have a signal-to-noise-ratio $S/N \geq 3$ in all the three bands of selection 
(3.4, 4.6 and 12 $\mu$m: W1, W2 and W3 from now) and are unaffected by known artifacts (cc$\_$flags ==`0000'). This implied the same criteria as used 
by \cite{2015ApJS..221...12S} albeit with a more relaxed S/N requirement of 3 instead of 5. \cite{2015ApJS..221...12S} adopted a criterion of $S/N \geq$ 5 in order
to minimise the contamination from star-forming galaxies. However, the presence of a corresponding X-ray counterpart of an ALLWISE source makes this less likely.
The relaxed S/N criterion enabled us to increase our sample of detected AGN, while our choice of mid-infrared colour selection is well suited to also find 
AGN behind highly obscured regions of the SMC.

\subsection{HMQ/MILLIQUAS catalogues}
In order to find more sources for the reference AGN sample that were either not detected in the ALLWISE data, or did not have the colours matching the \citet[]{2012MNRAS.426.3271M}
criterion, we selected additional high confidence sources with
known spectroscopic redshifts
from the HMQ and the MILLIQUAS catalogue within the area of the \xmm\ SMC survey. 
 This resulted in a total of 2753 sources from the  ALLWISE data and the HMQ/MILLIQUAS, out of which 2587 (94\%) are from the ALLWISE data. 

\subsection{Correlation of the \xmm\ and ALLWISE/HMQ/MILLIQUAS catalogues}
\begin{figure}
\centering
\resizebox{0.8\hsize}{!}{\includegraphics[]{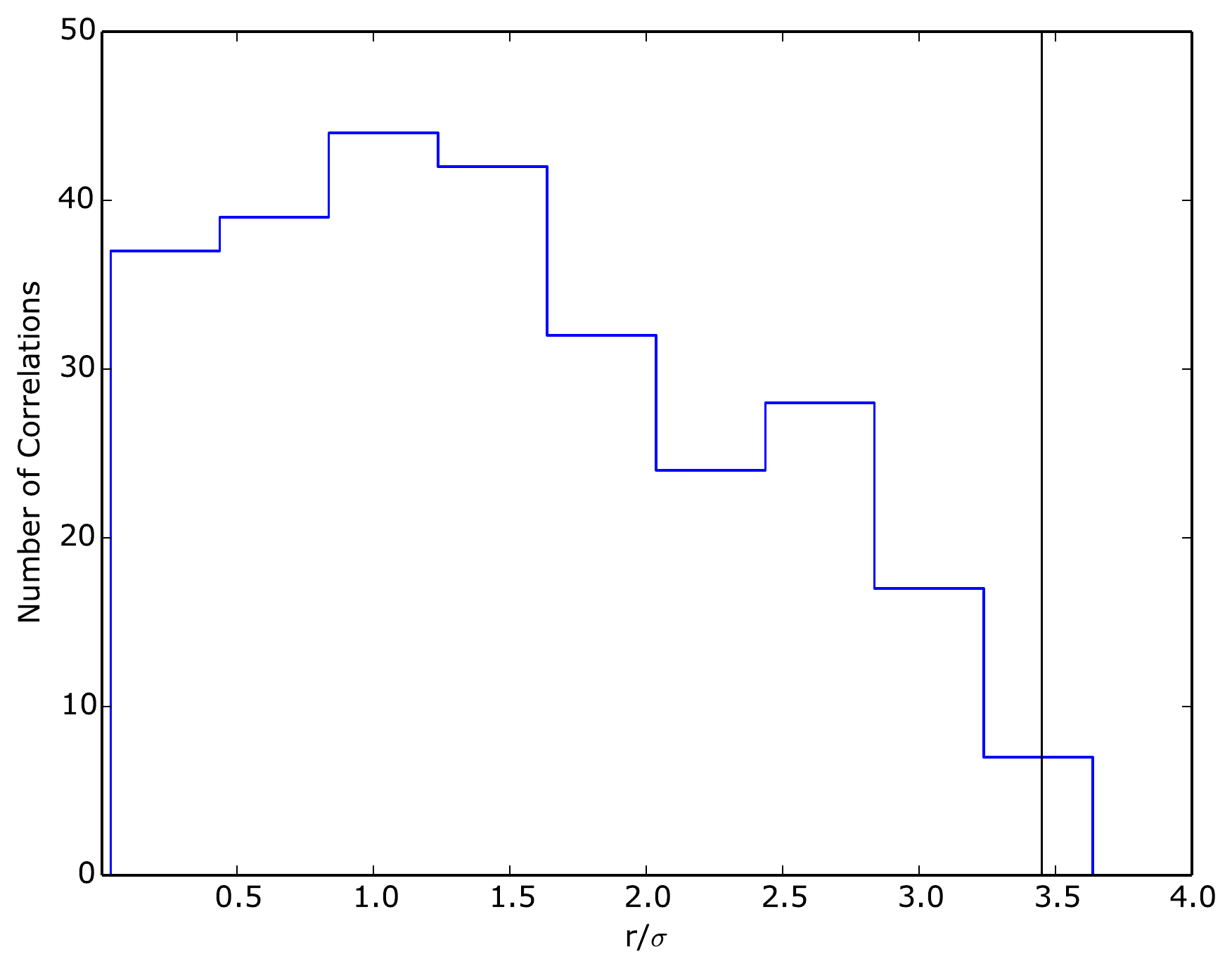}}
\caption{Distribution of uncertainty-normalised angular separations for 276 \xmm\ correlations with the total AGN sample of 2753 sources.
For real counterparts, we expect a Rayleigh distribution. The vertical line indicates the 99.7\% completeness.}
\label{rsigma}
\end{figure}

In order to identify the AGN behind the SMC from the survey data, we cross-correlated the X-ray and the ALLWISE/HMQ/MILLIQUAS samples described in the previous subsections.
For the cross-match we used TOPCAT v 4.2-1\footnote{http://www.star.bris.ac.uk/~mbt/topcat/}, and adopted the following selection methods to optimise the purity of the resulting sample.
Sources with uncertain X-ray coordinates can result in more than one match, which is therefore insecure. 
In the following, we only used X-ray sources with a 1$\sigma$ position uncertainty of $\sigma_{\rm X}<2\rlap{.}\arcsec5$ (4311 out of 4449 sources). We further regarded all correlations 
with an angular separation of 
      \begin{eqnarray*} 
       r \leq 3.439 \times  \sqrt{\sigma_{\rm X}^2+\sigma_{\rm MIR/HMQ/MILLIQUAS}^2} = 3.439\sigma.
      \end{eqnarray*}
For a Rayleigh distribution this corresponds to a 99.7\% completeness. The typical position accuracy of an X-ray source is $1\rlap{.}\arcsec2$. The ALLWISE position uncertainties are $< 1\arcsec$ for all 
sources and better than $0\rlap{.}\arcsec15$ for high S/N sources \citep{2010AJ....140.1868W}.
The coordinate uncertainties of the HMQ/MILLIQUAS sample is $\sim 0\rlap{.}\arcsec2$. Fig.~\ref{rsigma}, shows the distribution of the uncertainty normalised angular separation
between the correlated sources after applying the criteria described above.
      
To estimate the probability of chance coincidence, we cross-correlated the X-ray sample with the total sample of AGN by shifting 
the coordinates of the latter by large offsets (3\arcmin\, in either R.A. or Dec.).
We estimate the probability of chance coincidence to be 0.003\%.

\begin{table*}
\caption[]{AGN behind the SMC selected in this work.}
\begin{center}
\begin{tabular}{lll}
\hline\hline\noalign{\smallskip}
\multicolumn{1}{l}{Reference} &
\multicolumn{1}{l}{Number of sources} &
\multicolumn{1}{l}{Comments} \\
\noalign{\smallskip}\hline\noalign{\smallskip}
{\citet{2015ApJS..221...12S}} & 137 & 29 also included in HMQ/MILLIQUAS \\
HMQ/MILLIQUAS & 58 & None in ALLWISE \\
New candidates  &  81 (Table \ref{tab:new}) & selected by relaxing the S/N criterion in the ALLWISE data \\
  &   & 59 catalogued as AGN candidates for the first time \\
\noalign{\smallskip}\hline
\end{tabular}
\end{center}
\label{tab:catref}
\end{table*}

\section{Results and discussion}
\label{sec:analyses}
\subsection{The X-ray selected sample: characterisation}

\begin{figure*}
\centering
\resizebox{0.98\hsize}{!}{\includegraphics[]{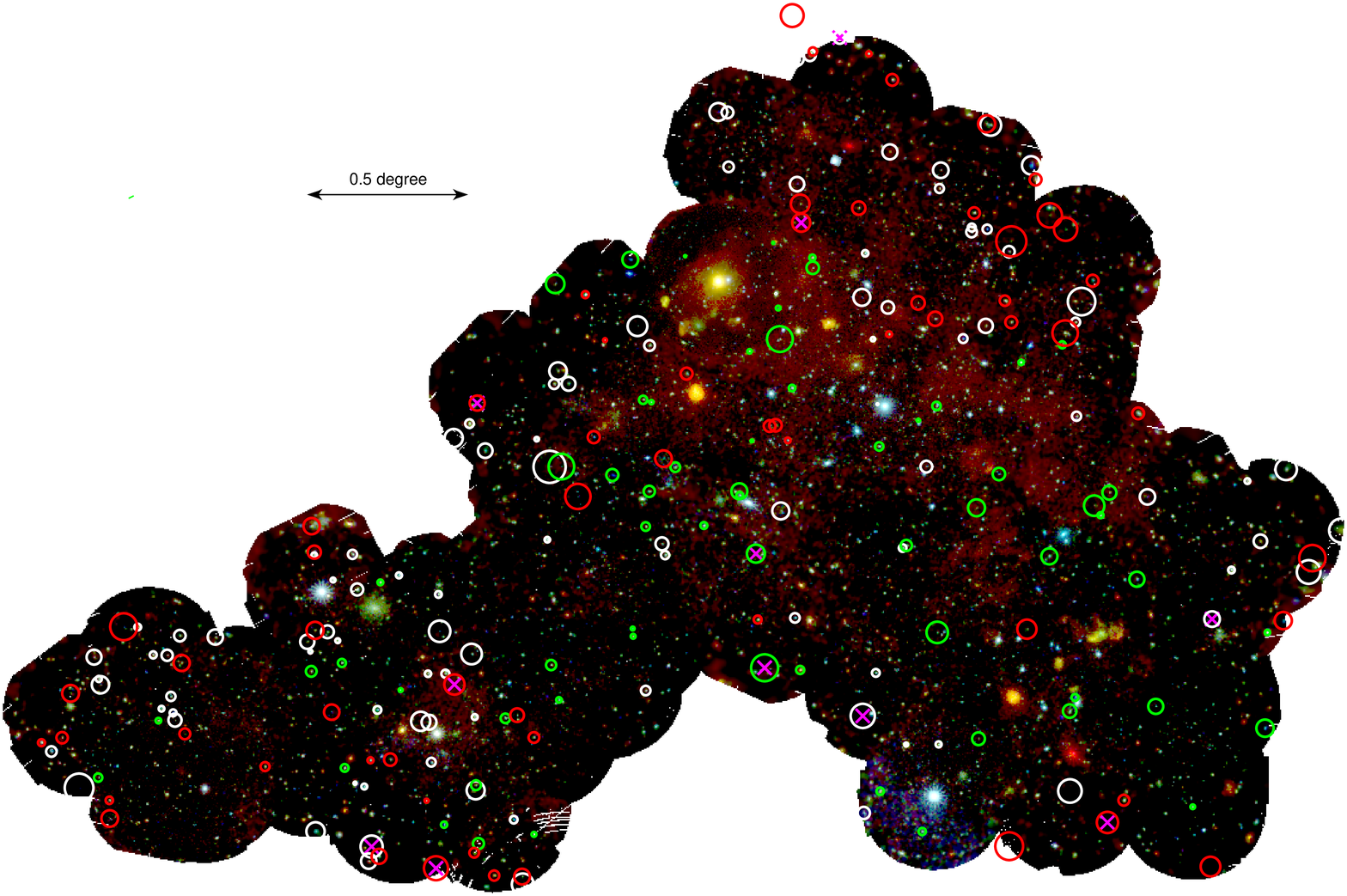}}
\caption{Mosaic image of all \xmm\ observations used in the updated SMC survey. Red/green/blue: 0.2--1.0 /
1.0--2.0 / 2.0--4.5 keV intensities. White circles denote the sources selected from \cite{2015ApJS..221...12S}, green circles
those from the HMQ and MILLIQUAS survey \citep{2015PASA...32...10F,2017yCat.7277....0F}, and red circles mark the 
81 new AGN candidates which we identified. The magenta crosses indicate the candidates for obscured AGN.
The circle radius is scaled with the uncertainty on the X-ray positions.}
\label{spatial}
\end{figure*}

We found a total of 276 X-ray sources uniquely correlating with the selected sample of AGN (Table \ref{tacat}). Out of them 218 in total
were selected using the ALLWISE selection criteria, with 81 new AGN candidates
identified from this work, and 137 already classified as AGN by \cite{2015ApJS..221...12S} (Table~\ref{tab:catref}). The rest (58) were selected exclusively from the MILLIQUAS and the HMQ catalogue. The spatial distribution of the sample (Fig.~\ref{spatial}), shows that the 
identified AGN are scattered uniformly within the \xmm\ survey area with no particular spatial dependence on the selection criteria (ALLWISE/HMQ/MILLIQUAS) used for the sample. 

\subsubsection{Redshift}
From the 276 sources, 90 have their redshifts ($z$) determined. Their distribution is shown in Fig.~\ref{distribution}, right. 
The median $z$ of the sample is 1.06, the lowest is 0.07 (classified as a Seyfert galaxy 6dFGS gJ005356.2--703804) and the highest is 2.878 (classified as quasar SMC J010127.75$-$721306.2). This indicates that we detect sources with a wide range of $z$ from nearby Seyfert galaxies to distant quasars.
We detected 14 objects with high redshifts ($z \geq$ 2). Although most of the sample was selected by correlating the X-ray source list with the ALLWISE 
mid-infrared two colour selection, only 29 of the objects with known z are from the ALLWISE sample. This is consistent with the fact that out of 
the 2587 sources identified within the 
area of the \xmm\, survey that satisfied the criterion of \cite{2012MNRAS.426.3271M}, only 73 contain redshift information.
The rest was selected from the MILLIQUAS and the HMQ catalogue.
Moreover, all the sources with  $z \geq$ 2 were also selected from the MILLIQUAS and the HMQ catalogue.
We verified that the high $z$ sources were distributed randomly within the survey area of the SMC, further indicating that there is no dependence of the selection criteria
used for identification of the sources on the area of the survey.

\subsubsection{Flux and luminosity}

\begin{figure*}
\centering
\resizebox{0.45\hsize}{!}{\includegraphics[]{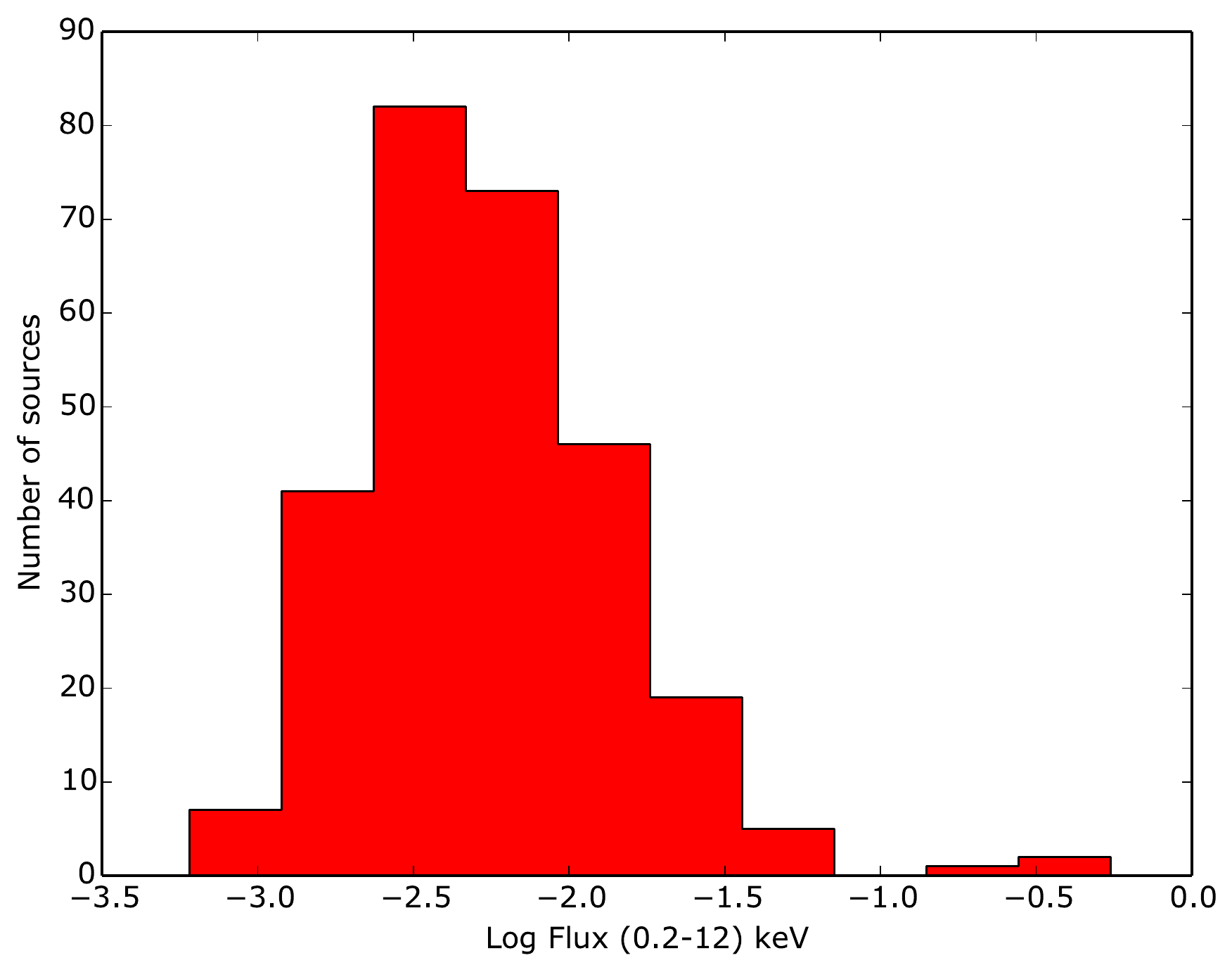}}
\resizebox{0.40\hsize}{!}{\includegraphics[]{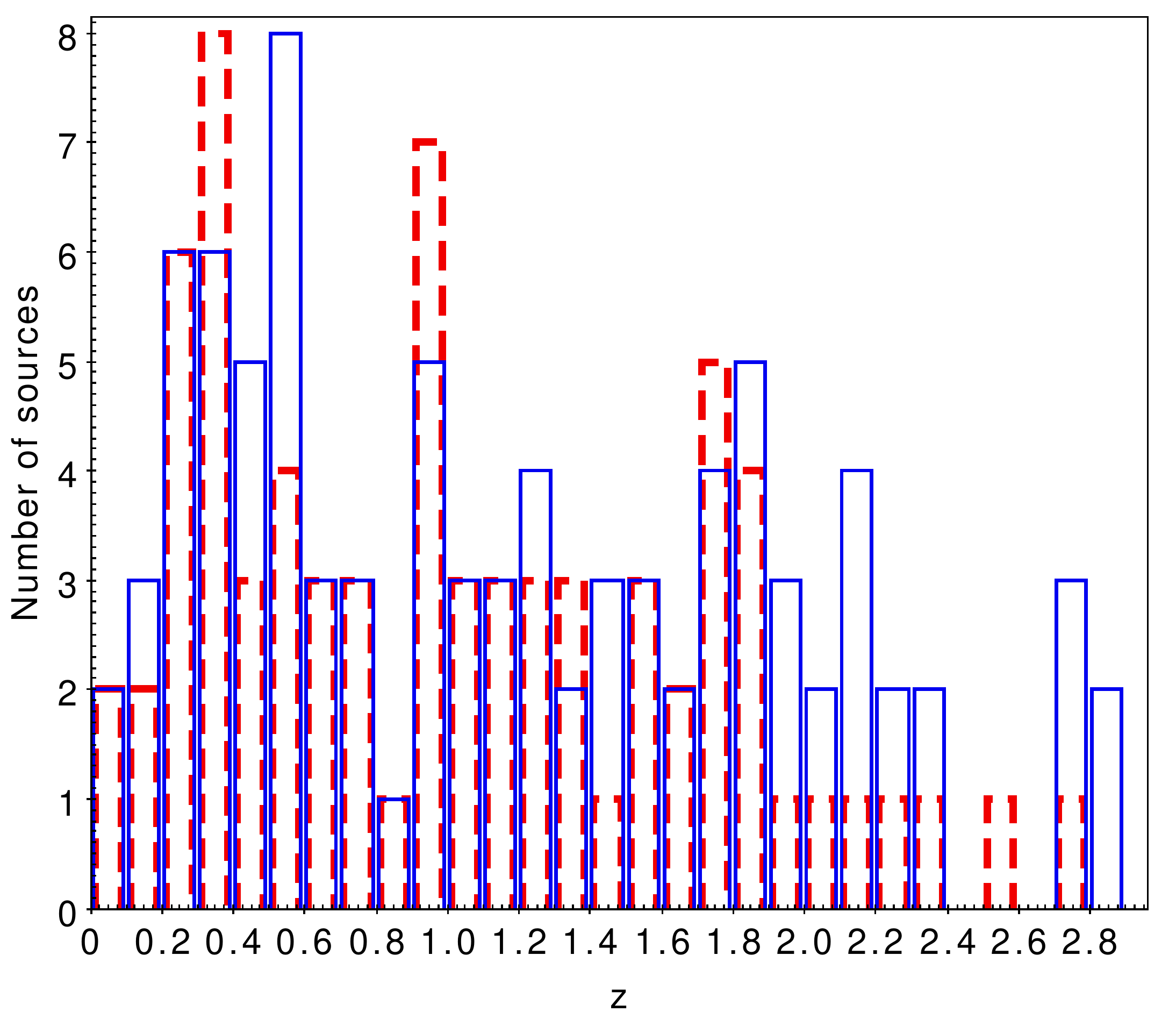}}
\caption{{\it Left}: X-ray flux (in units of \oergcm{-11} in the energy range of 0.2--12 keV) distribution of the sample; 
{\it Right}: Redshift distribution of the sample where known (in red) and redshift distribution of all AGN with known $z$ (in blue) within the \xmm\ SMC survey
from \citet[][]{2015ApJS..221...12S}.}
\label{distribution}
\end{figure*}

\begin{figure*}
\centering
\resizebox{0.45\hsize}{!}{\includegraphics[]{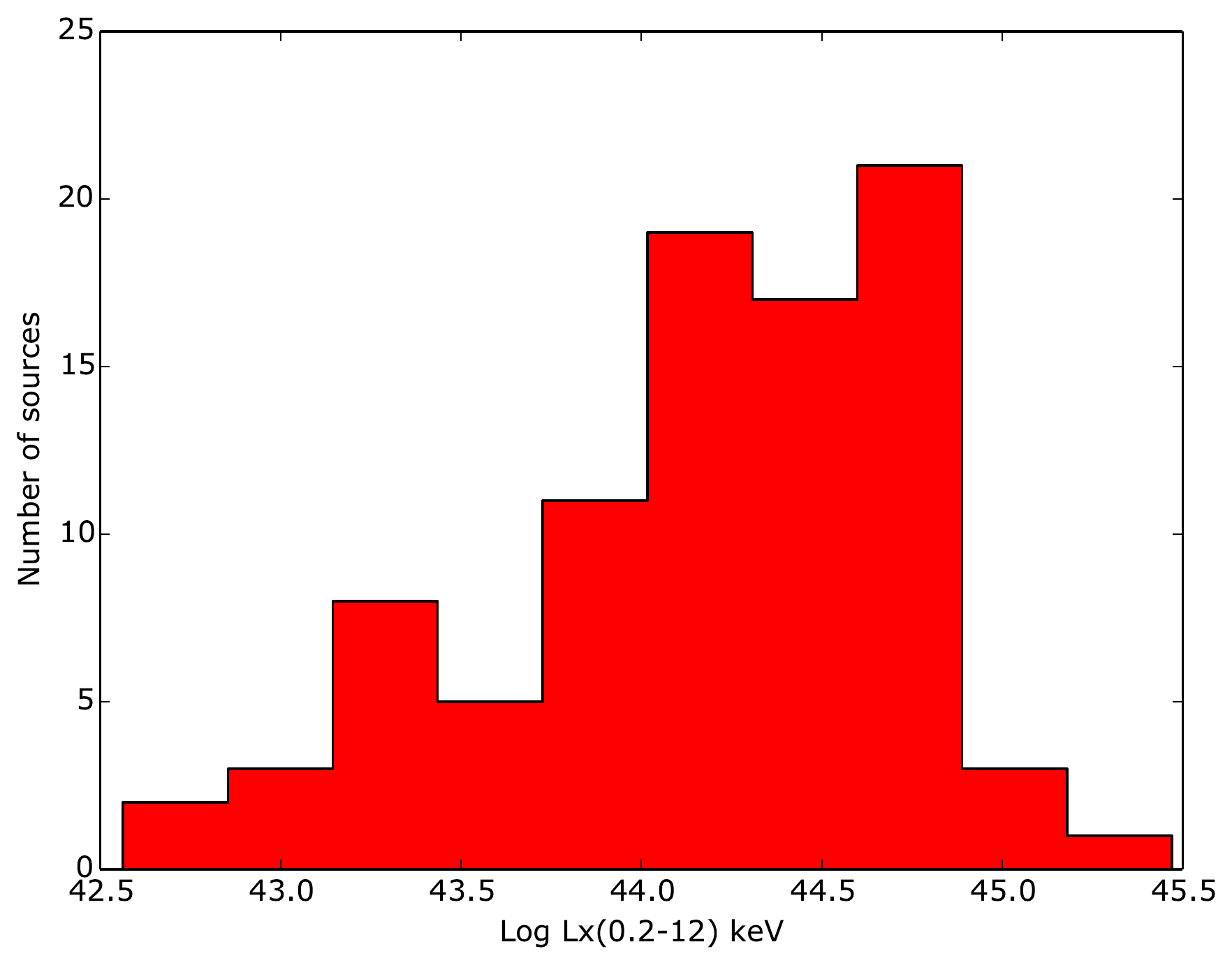}}
\resizebox{0.45\hsize}{!}{\includegraphics[]{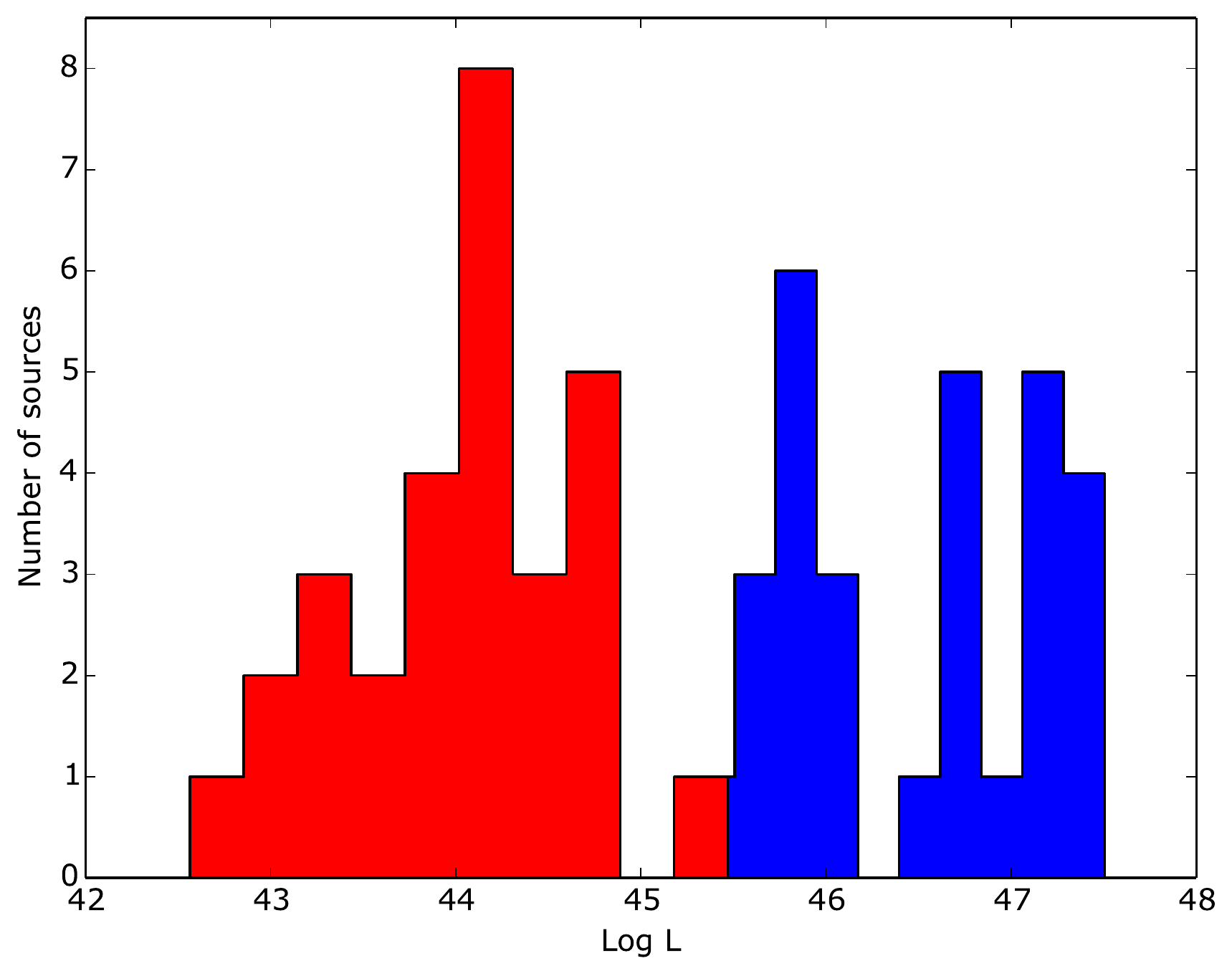}}
\caption{{\it Left}: X-ray luminosity (in units of erg s$^{-1}$) distribution of the sub-sample with known redshift; 
{\it Right}: Comparison between X-ray (red) and mid-infrared (blue) luminosity distribution from a smaller set where
both the X-ray and mid-infrared luminosities can be determined.}
\label{lx-distribution}
\end{figure*}

We calculated the average X-ray flux for each source of the sample (0.2--12 keV) by converting the count rates using conversion factors taken from \cite{2013A&A...558A...3S}. 
This assumes a power-law model with a photon index of 1.7 and a photo-electric foreground absorption by the Galaxy of  $6\times10^{20}$ cm$^{-2}$ \citep[average for the SMC 
main field in the \Hone\, map of][]{1990ARA&A..28..215D}.
The flux of the sample has a median at 7.0\ergcm{-14} (Fig.~\ref{distribution}, left). 
The minimum flux detected in the sample is 6.0\ergcm{-15}, which is near the sensitivity limit of the survey. 
The highest flux detected is 5.0\ergcm{-12} from a nearby object ($z=0.074$). 

For the sources with known $z$, we determined the rest-frame X-ray luminosities (0.2--12 keV). The X-ray luminosity distribution is shown in Fig.~\ref{lx-distribution}, left.
The median of the sample is 1.9\ergs{44} and the minimum luminosity determined is 3.7\ergs{42}. The lower limit of our sample 
is compatible with the value at which the AGN luminosity dominates over the host galaxy emission and can be selected effectively using AGN colour selection criteria \citep[][and 
references therein]{2015ApJ...807..129S}. We note however, that the luminosities were calculated using the flux obtained from converting the count rates alone, and were not 
corrected for obscuration effects due to local or the
line-of-sight absorption. Hence it does not reflect the intrinsic luminosity of the sample. In order to obtain a robust estimation of the true luminosity,
we determined the $6 \mu$m rest-frame luminosity for the 29 sources containing information on the magnitudes in the four ALLWISE bands \citep[(3.4, 4.6, 12 and 22 $\mu$m,][]{2017MNRAS.468.3042M}. 
The choice of the wavelength was motivated by the fact that 6 $\mu$m provides an unbiased estimate of the bolometric luminosity representing the torus luminosity.
Further, the contribution from star-formation is negligible at this wavelength \citep{2017MNRAS.468.3042M}. 
Figure~\ref{lx-distribution}, right shows the distribution of the infrared luminosity (in red) and the X-ray luminosity (in blue) for the sample. 
The median of the infrared-luminosity ($\nu L_{\nu}$) distribution is 4.7\ergs{46} and the highest determined value is 3.2\ergs{47} (indicative of a powerful quasar).

\begin{figure}
\centering
\resizebox{0.95\hsize}{!}{\includegraphics[clip=]{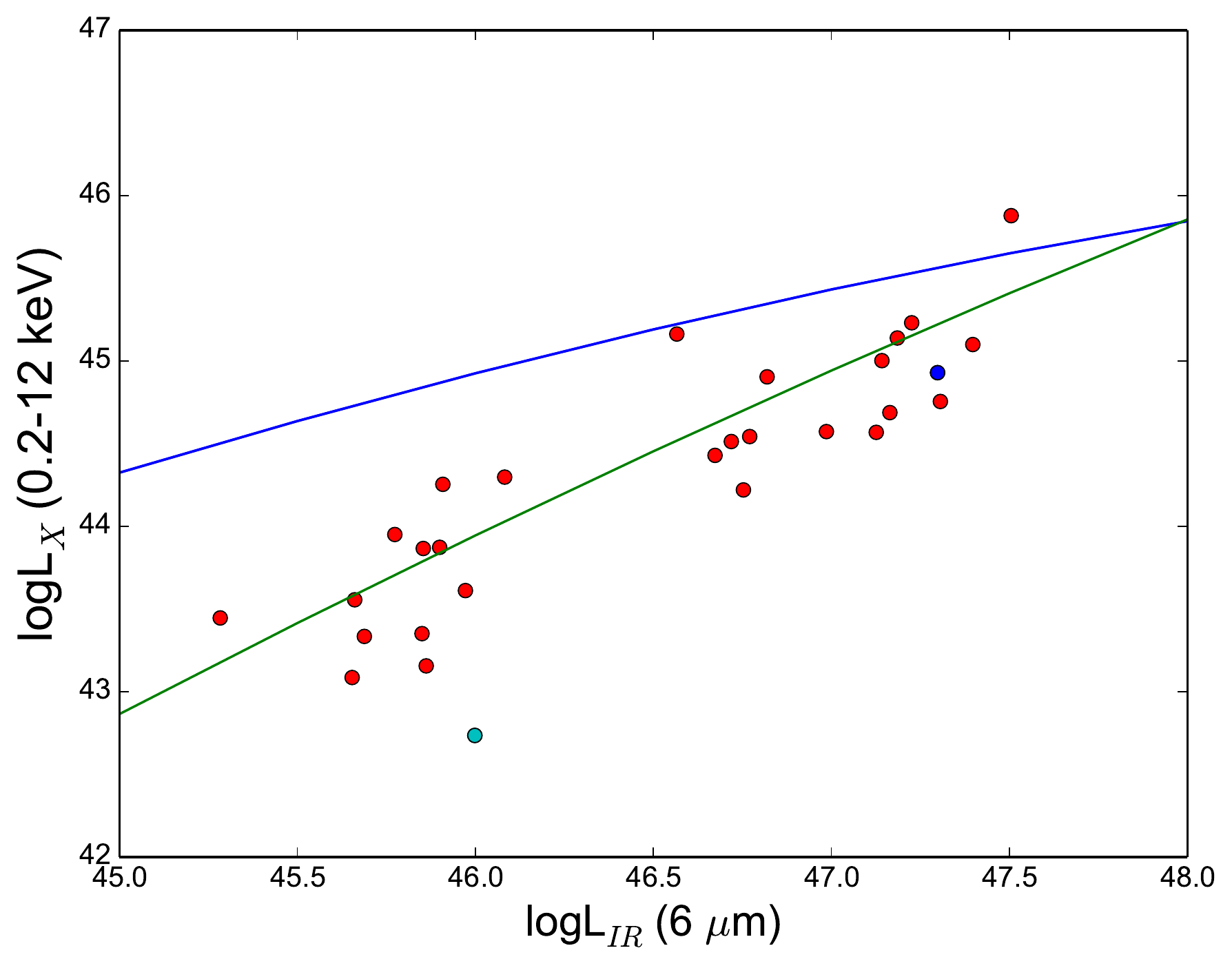}}
\caption{Comparison of X-ray (0.2--12 keV) and mid-infrared (6 $\mu$m) luminosity (in units of erg s$^{-1}$) shown in Fig.~\ref{lx-distribution}.
The blue line denotes the X-ray-mid-IR luminosity relation from \cite{2015ApJ...807..129S}. 
The blue dot marks the candidate for obscured
AGN identified by comparing the X-ray hardness-ratio of the sample. The cyan dot marks the candidate for obscured AGN identified 
from an unusually low $L_{x}/L_{IR}$ ratio. The green line denotes the relation derived from this work
for the current sample.}
\label{fzx}
\end{figure}

Figure~\ref{fzx} shows the comparison of the rest-frame X-ray (0.2--12 keV, not corrected for absorption) and the mid-infrared (6 $\mu m$) luminosity 
of this sample. Previous works  in this regard have demonstrated how the luminosities are correlated
\citep{2009A&A...502..457G,2009ApJ...693..447F,2009A&A...498...67L}, including \cite{2015ApJ...807..129S} who formulated the global X-ray to mid-IR relation of AGN which is
appropriate for a large range of luminosities from Seyfert galaxies to luminous quasars. Figure~\ref{fzx} shows that the brightest objects in mid-IR are also bright in X-rays. 
To quantify the correlation, we determined the Kendall's $\tau$ rank correlation coefficient to be 0.70.  Artificial correlations may be introduced between luminosities due 
to the effect of redshift in a flux-limited sample. In order to take this into account, we determined the partial Kendall's $\tau$ rank correlation coefficient to
be 0.44, indicating a weaker
intrinsic correlation between the X-ray and mid-IR luminosities.

We also compare our results with the X-ray to mid-IR relation of AGN. The X-ray luminosities determined in this work are systematically lower than predicted by the relation. The bias indicates the presence of significant
X-ray obscuration. This is because the formulation of \cite{2015ApJ...807..129S} is based on absorption corrected luminosities in the energy band of 2--10 keV.
The X-ray luminosities reported in this work on the other hand are obtained from converting count rates to average flux (0.2--12 keV), and accounting only for
 a photo-electric foreground absorption by the Galaxy of $6\times10^{20}$ cm$^{-2}$.
 As the AGN are distributed uniformly within the \xmm\, survey area (Fig.~\ref{spatial}), an additional obscuration of \nh\ $= 2\times10^{21}$ cm$^{-2}$ through the depth of the SMC \citep[as is typically found in the wing region of the SMC;][]{1999MNRAS.302..417S} corresponds to an increase in the absorption-corrected luminosity of
 $\sim$ 2\% in the 2--10 keV energy range and 15\% in the 0.2--12 keV range.
Similar factors corresponding to \nh $= 10^{22}$ cm$^{-2}$  \citep[as is typically found in the bar region of the SMC;][]{1999MNRAS.302..417S} are 8\% and 30\% respectively. We performed a least-square polynomial fit to 
the data and derived a slightly
different relation (not corrected for absorption) than \cite{2015ApJ...807..129S}:
$$\log L_{x}({\rm 0.2-12~keV}) = 37.711 + 1.455 x - 0.042 x^2,$$ where L$_{x}$ is in units of erg s$^{-1}$ and $x \equiv$ $\log (\nu L_{\nu}/10^{41}\,$ erg s$^{-1}$).
A change in slope of our obtained relation as compared to the global formulation of \cite{2015ApJ...807..129S} indicates that the intrinsic absorption of the AGN is an important contributing factor to the total absorption component.
As the AGN intrinsic absorption is redshift dependent, it would have a higher effect on nearby, low luminosity Seyfert galaxies as compared to luminous and distant quasars
in the observers reference frame.

\subsubsection{X-ray Spectral characteristics}

\begin{figure}
\centering
\resizebox{0.9\hsize}{!}{\includegraphics[]{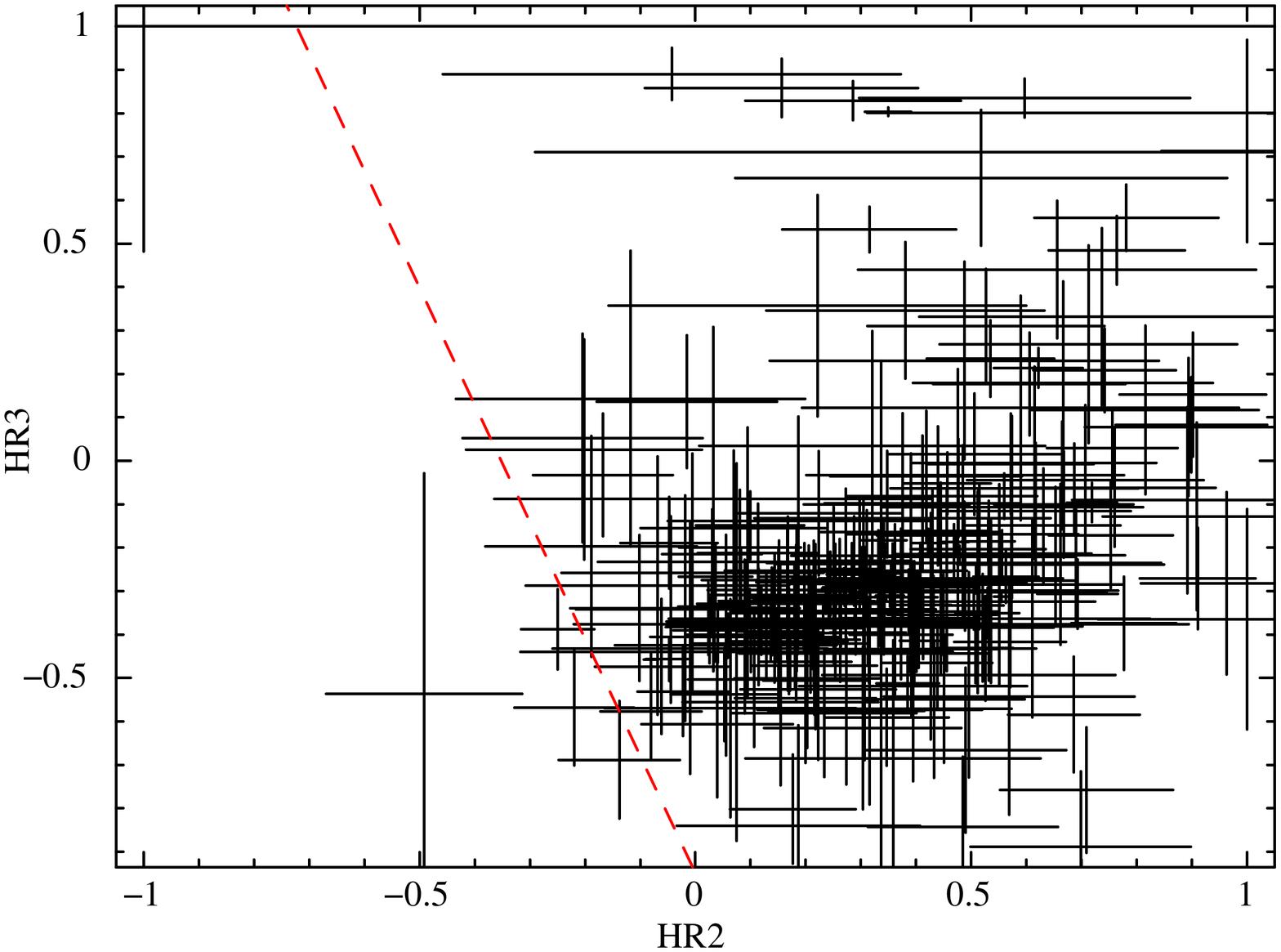}}
\resizebox{0.9\hsize}{!}{\includegraphics[]{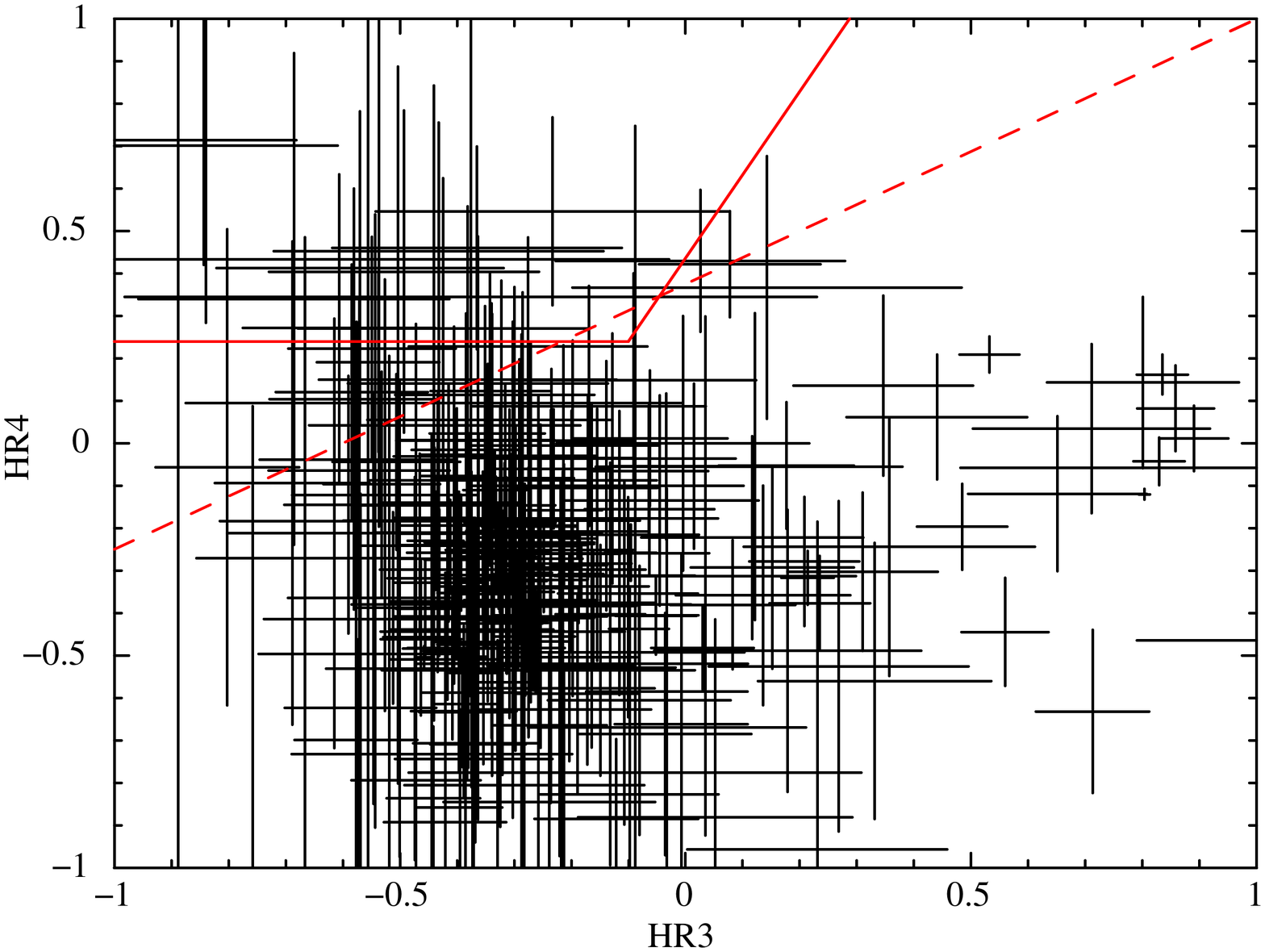}}
\caption{{\it Top}: Hardness ratio diagram for the AGN sample. The dashed line defines the spectral classification criterion in HR3 vs. HR2 by \citet{2013A&A...558A...3S} to identify AGN. 
{\it Bottom}: Hardness ratio diagram as the top figure, albeit in different energy bands. The solid wedge and the dashed line defines the selection criteria for highly obscured AGN in \citet{2012MNRAS.422.1166B}.}
\label{hr2hr3}
\end{figure}

The X-ray spectra of an AGN can typically be described by a power law with photon index of $\Gamma\sim 1.75$ \citep[e.g.][]{2006A&A...451..457T} 
and therefore show a hard X-ray spectrum compared to Galactic stars.
As shown by \citet{2013A&A...558A...3S}, hard X-ray sources can be selected efficiently using the hardness-ratio relation
\begin{eqnarray*} 
8 HR_2 + 3 HR_3  & > & -3
\end{eqnarray*}
or sources with $HR_2>0$ if $HR_3$ is not defined, due to lack of statistics.
Here, hardness ratios are derived by comparing count rates in neighbouring \xmm\ energy bands: $HR_i=(R_{i+1}-R_{i})/(R_{i+1}+R_{i})$, 
where $i$=1 for 0.2--0.5 keV, $i$=2 for 0.5--1.0 keV, $i$=3 for 1.0--2.0 keV, 
$i$=4 for 2.0--4.5 keV and $i$=5 for 4.5--12.0 keV.
 We extracted this information for the AGN sample and plotted the HR2--HR3 hardness--hardness diagram as shown
in Fig.~\ref{hr2hr3}. We found that almost all the sources (except a few with very large error bars) lie within the specified region for AGN
in the HR2--HR3 plane prescribed in \cite{2013A&A...558A...3S}. This further validated the robustness of our selection.
    
The mid-infrared based surveys have the potential to find the most obscured and even Compton-thick AGN (\nh\ $\ge 10^{24}$ cm$^{-2}$), and thus helps in 
identifying X-ray selected heavily obscured AGN \citep[][and references therein]{2015ApJS..221...12S}. 
This motivated us to look for obscured AGN within the sample. \cite{2012MNRAS.422.1166B} showed that heavily obscured sources can be well separated 
in X-rays
by being hard in the band HR2 ($\equiv$ to HR4 band in our definition) and soft in HR1 ($\equiv$ to HR3 band in our definition). Using equation (1) and (2) from 
\cite{2012MNRAS.422.1166B}, we found that 13 sources satisfy the criteria of expected \nh\ $\ge 10^{24}$ cm$^{-2}$ (Compton-thick) and 20 sources
satisfy the criteria of expected \nh\ $\ge 10^{23}$ cm$^{-2}$ (see Fig.~\ref{hr2hr3} and Table\,\ref{taobscured}). 
Alternatively, an unusually low $L_{x}/L_{IR}$ ratio has been used to test the Compton-thick nature of candidates in nearby ($z< 0.1$) AGN \citep{2012A&A...542A..46S,2014MNRAS.438..494R}. 
With this approach we found one more candidate by looking for extreme outliers in the X-ray to mid-IR luminosity comparison 
(source marked cyan in Fig.~\ref{fzx}).
All the 21 sources were distributed evenly within the survey area, except the bar region of the SMC, where the absorption within the SMC is the highest (see Fig.~\ref{spatial}).
This is further indicative of the fact that the sources are intrinsically absorbed. The high absorption in the bar region also makes the X-ray detection of the obscured AGN difficult.
Interestingly, for three of these sources  $z \sim 2$, which is indicative of distant objects. 
The possible luminous, distant and obscured quasar candidates are rare objects, and their identification is crucial to reproduce the shape of the cosmic X-ray background \citep[see for instance][]{2007A&A...463...79G}.
These sources therefore deserve attention for further follow-up studies.
The sample of possibly obscured AGN also contains 8 newly identified candidates described in section~\ref{sec:new}. The fraction of Compton-thick
 AGN candidates constitute $\sim$ 5\% of the total sample and is roughly consistent with the fraction detected in the \xmm\, survey of the COSMOS field \citep{2007ApJS..172...29H} 
 and the Lockman hole \citep{2001A&A...365L..45H}.
  
To confirm their nature, it would be ideal to investigate the X-ray spectra of the sample to constrain the \nh\,, determine the unabsorbed X-ray luminosities of the sources, or look for 
signatures of obscured AGN such as a convex-shaped X-ray spectrum indicative of very high  \nh\ or a Fe-K emission line of high equivalent width \citep[see e.g.][]{2015ApJ...814...11T}.  
All the sources were however either detected off-axis and/or are X-ray faint. 
Therefore, the statistical quality of the source spectra were inadequate for performing a meaningful analysis. 
The identified sources are nevertheless promising targets for follow-up studies of distant obscured quasars.



\subsection{Correlation with the SMC X-ray point source catalogue from 2013}
\label{sec:new}

We correlated our sample with the SMC point source catalogue of \cite{2013A&A...558A...3S} which is based on a subset of the observations used for this work. 
We found that 201 of our 276 sources are included in that catalogue. Forty of them were already classified as AGN. The remaining sources were either unclassified or candidate AGN. 
Our work further validates the candidate sources as real AGN. Only 8 out of the selected sources were included in the sample of 88 AGN with radio associations \citep{2013A&A...558A.101S}. 
We note that although 12 more out of the 88 sources had a counterpart in the ALLWISE catalogue, their mid-infrared colours did not meet 
the selection criterion of \cite{2012MNRAS.426.3271M}.
The very small overlap between the sample selected from radio associations, and the present selection methods indicate that they are perhaps sensitive to different 
populations of AGN (blazars and radio galaxies vs. Seyfert galaxies).

\subsection{Identification of near-infrared counterparts}

It is well known that the AGN are prominent in the near-infrared band, and their variability can be used to identify 
them \citep[e.g.][]{1987ApJ...321..233E}.

These two considerations prompted us to cross-identify our objects
with the point source catalogue of the VISTA \citep[Visual and Infrared 
Survey Telescope for Astronomy;][]{2006Msngr.126...41E} Survey of 
the MC system \citep[VMC;][]{2011A&A...527A.116C}. 
VMC is an ESO public survey, carried out with the wide-field near-infrared 
camera VIRCAM \citep[VISTA InfraRed CAMera;][]{2006SPIE.6269E..0XD}, 
mounted at the European Southern Observatory's (ESO) VISTA 4.1 m 
telescope. More information about the telescope, the data flow system
and the archive access to the data is given in \cite{2004SPIE.5493..411I,2004SPIE.5493..401E,2006Msngr.126...41E,2018MNRAS.474.5459G,2012A&A...548A.119C}
and \citet{2011A&A...527A.116C} provides a detailed description of 
the VMC's observing strategy. In short, the VMC covers 170\,deg$^2$ 
around the MCs, Bridge and Stream, down to 
$K_\mathrm{s}$=20.3\,mag ($S/N \sim$10; Vega system) in three epochs 
in the $Y$ and $J$, and in 12 epochs (each of which consists of two or more individual
pawprint observations) in the $K_\mathrm{s}$ band, 
spread over a year or longer. 

\begin{figure}
\centering
\resizebox{\hsize}{!}{\includegraphics[]{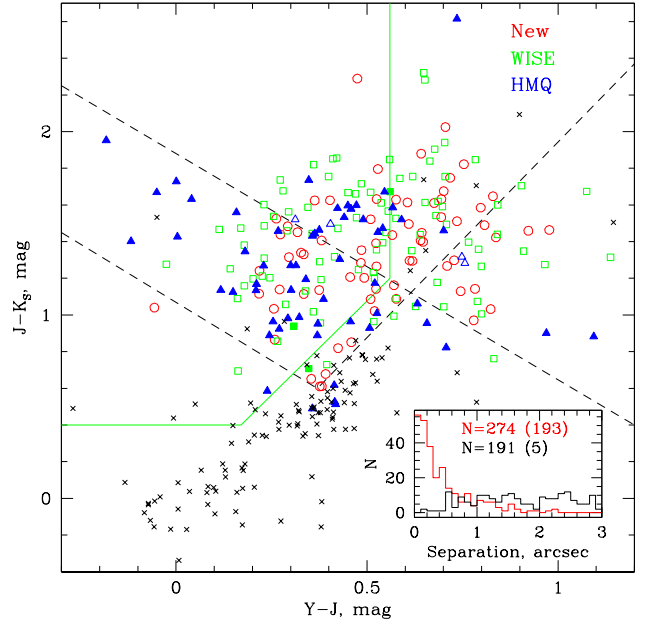}}
\caption{VMC colour--colour diagram. 
The lines follow Cioni et al. (2013) and show regions where known quasars (upper left of the dashed black lines) and planetary nebulae (left of the green wedge) were found. 
Our objects are marked with red circles if they are newly identified, green squares if they were identified in the WISE data,
or blue triangles if they were present in the HMQ/MILLIQUAS. The symbols are open if there has been no spectroscopic 
follow up, and solid if the objects have known redshifts from spectra. The reference sample is plotted with black crosses.
The inset shows histograms of the angular separations between the objects and their counterparts for our sample (red) and
for the reference sample (black). }
\label{cmd}
\end{figure}

At first we searched for the closest matches within 3$\arcsec$ for our 276 sources 
with the VMC obtained until March 2017 and identified 274 matches.
To evaluate the probability for random coincidence we repeated the 
search after moving the source coordinates 3$\arcmin$ to the east, and
found 191 matches -- these constitute a reference sample. 
The colour--colour diagram of the infrared counterparts with detections 
in all three bands is shown in Fig.\,\ref{cmd}. 
The inset shows histograms of the angular separation
for the closest match for both the main and reference samples.
Note that in many cases, there is more than one counterpart
within 3\arcsec; those were not counted in the histogram
for the main sample, and the higher counts in the reference
histogram at separations $\geq$ 1\arcsec\ reflect the
existence of additional VMC sources near the position of
the objects in our sample.
Apparently, the cross-identifications with separations $\leq0\rlap{.}\arcsec5$
are secure, and the random coincidences dominate the larger separations.
Therefore, we adopt $0\rlap{.}\arcsec5$ as a final matching radius for the VMC cross correlations and the subsequent analysis.
This limit guarantees virtually zero contamination, as can be seen from the separation histogram (Fig.~\ref{cmd}, inset panel).
Another conclusion we draw from this diagram is that the reference 
sample consists mostly of stars, which is clear from the similarity of the locus it occupies with the stellar locus in Fig.~2 from 
\citet{2013A&A...549A..29C}. The location of the counterparts strongly supports the AGN nature.

\begin{figure}
\centering
\resizebox{0.9\hsize}{!}{\includegraphics[]{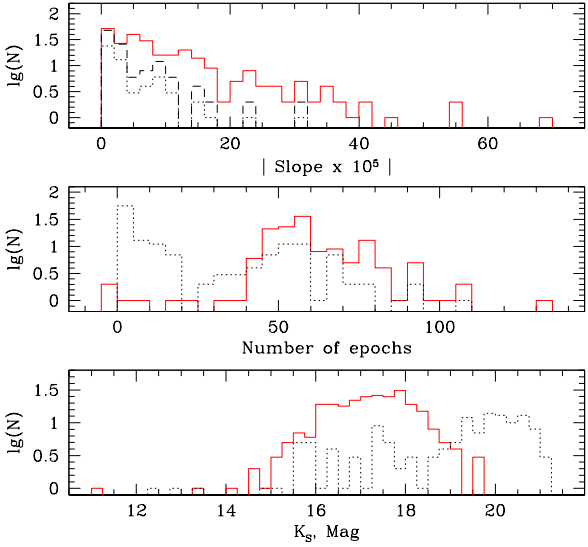}}
\caption{Properties of the VMC $K_S$ band light curves for the counterparts to our targets (red solid line) and of the light 
curves for the counterparts to the reference sample (black dotted line). The top panel shows histograms of the linear fits' slopes for 
objects with 15 or more epochs, so only reliable slopes are considered in the comparison. The black dashed line is the same 
as the black dotted, but scaled to match the level of the first bin. The middle panel shows histograms of the number of available $K_S$ 
epochs, and the bottom panel the $K_S$ band apparent luminosity functions averaged over epochs.}
\label{variability}
\end{figure}

Next, we extracted from the VMC catalogue the light curves for all the identified
counterparts.
For each light 
curve we calculated a linear fit of the $K_S$ magnitude versus 
time -- this is a simple parametrisation of the objects' 
variability, following \citet{2013A&A...549A..29C} and \citet{2016A&A...588A..93I}. The properties of the light curves of the target and reference 
samples are summarised in Fig.~\ref{variability}. Clearly, the identified counterparts are more 
variable than a randomly selected sample, as expected for a sample
dominated by AGN -- in good agreement with the results of \citet{2013A&A...549A..29C} and \citet{2016A&A...588A..93I}. Note that the difference in the slopes is more evident if we only 
consider objects observed with at least 15 epochs. Our experiments 
with the slopes indicated that a smaller number of observations 
produces less than reliable slopes.

The counterparts of our sample tend to be brighter than 
the counterparts of the reference sample. The latter conclusion 
accounts, at least partially, for the larger number of epochs of 
the counterparts to ``real'' objects than for the counterparts 
to the ``random'' objects. Finally, we note that no constraint was imposed on whether the VMC
matches are point-like or extended. This was motivated by the fact
that some of the confirmed low redshift quasars in \citet{2016A&A...588A..93I} were classified as extended, e.g. the VMC has enough spatial
resolution to see the host. Furthermore, background quasars can
be contaminated by imperfectly aligned foreground VMC stars, giving
the quasars an extended appearance.

\subsection{Multi-wavelength properties}
\begin{figure*}
\resizebox{0.48\hsize}{!}{\includegraphics[]{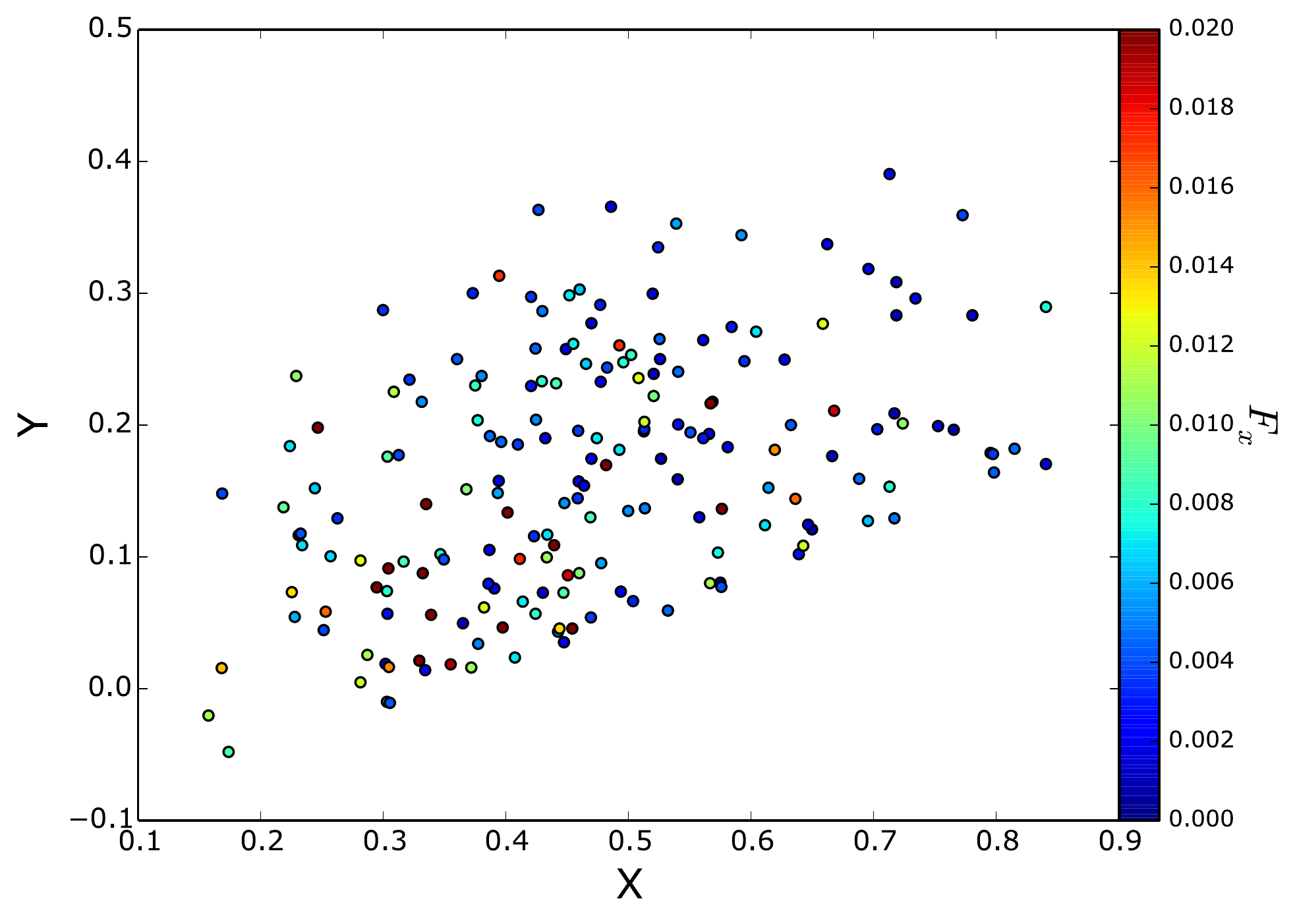}}
\resizebox{0.48\hsize}{!}{\includegraphics[]{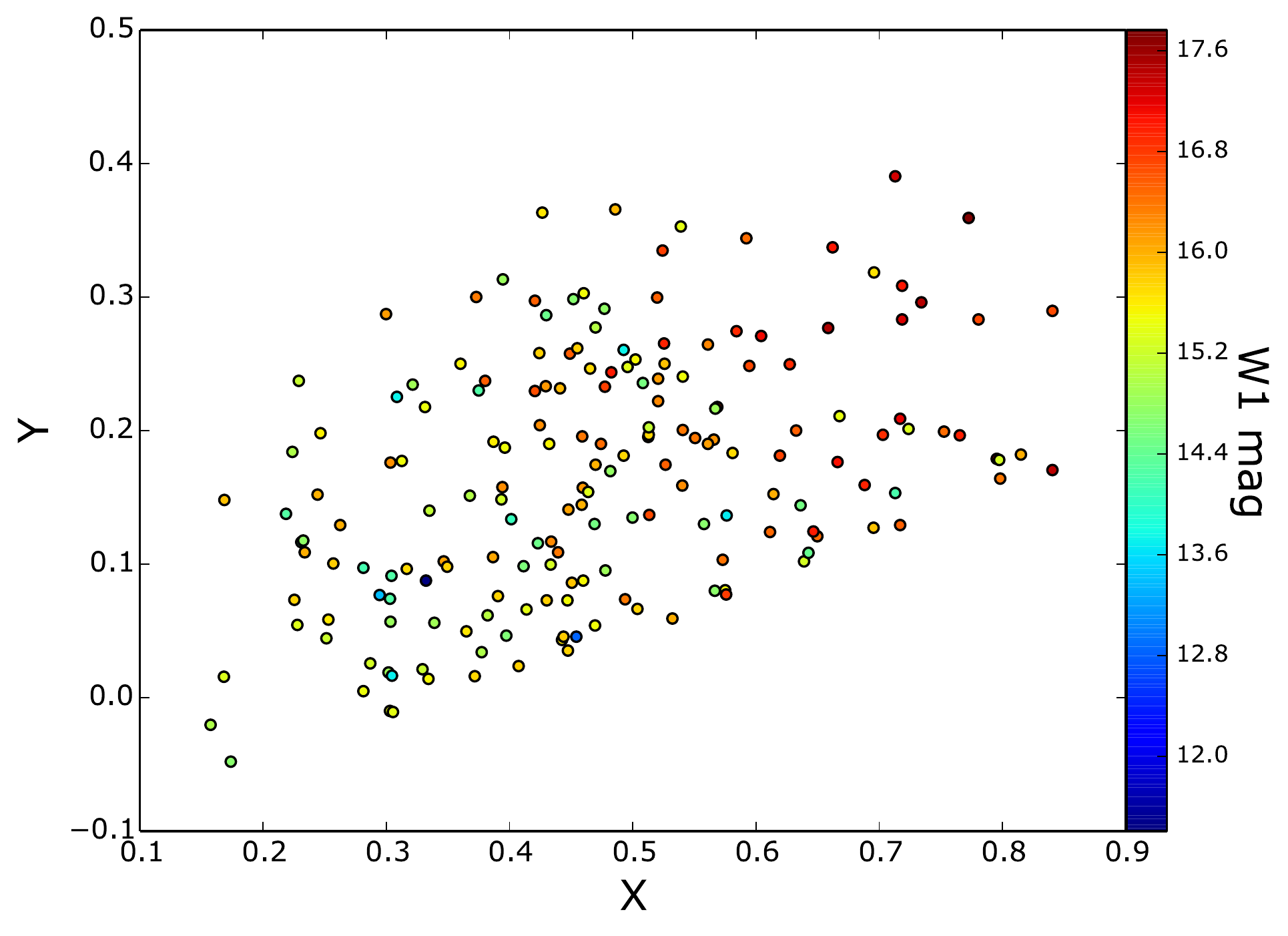}}

\caption{ Distribution of the sample in the ALLWISE two-colour selection plane \citep{2012MNRAS.426.3271M}.  $x \equiv \log \frac{(f12\mu m)}{(f4.6 \mu m)}$ and $y \equiv \log \frac{(f4.6\mu m)}{(f3.4 \mu m)}$. X-ray flux ($F_{x}$) is in units of \oergcm{-11} in the energy range of 0.2--12 keV. 
The two figures are representative of the distribution of the sample in the ALLWISE plane, and the variation with W2 and W3 magnitudes show a similar pattern.}
\label{allwise}
\end{figure*}

\begin{figure*}
\resizebox{0.48\hsize}{!}{\includegraphics[]{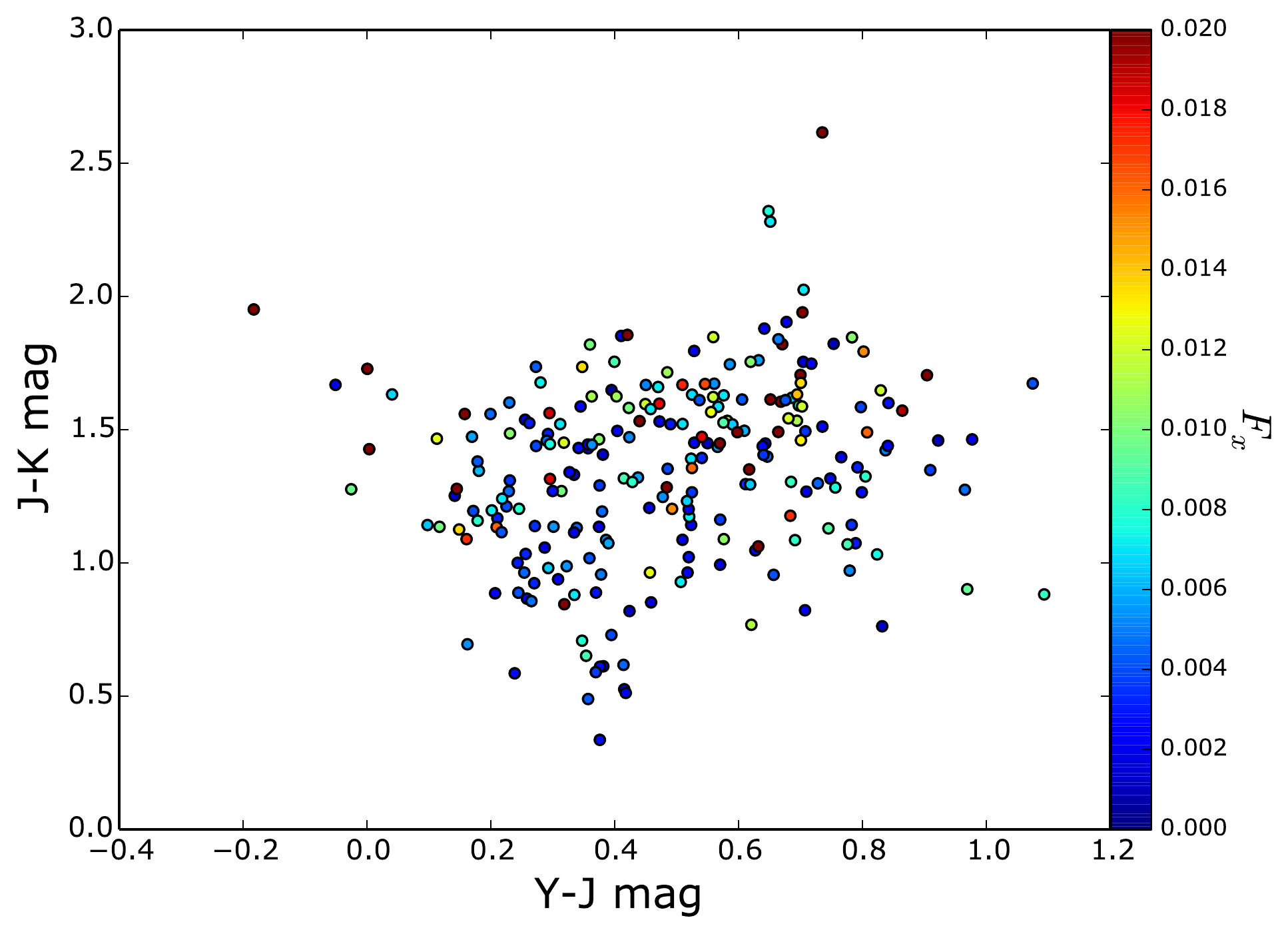}}
\resizebox{0.48\hsize}{!}{\includegraphics[]{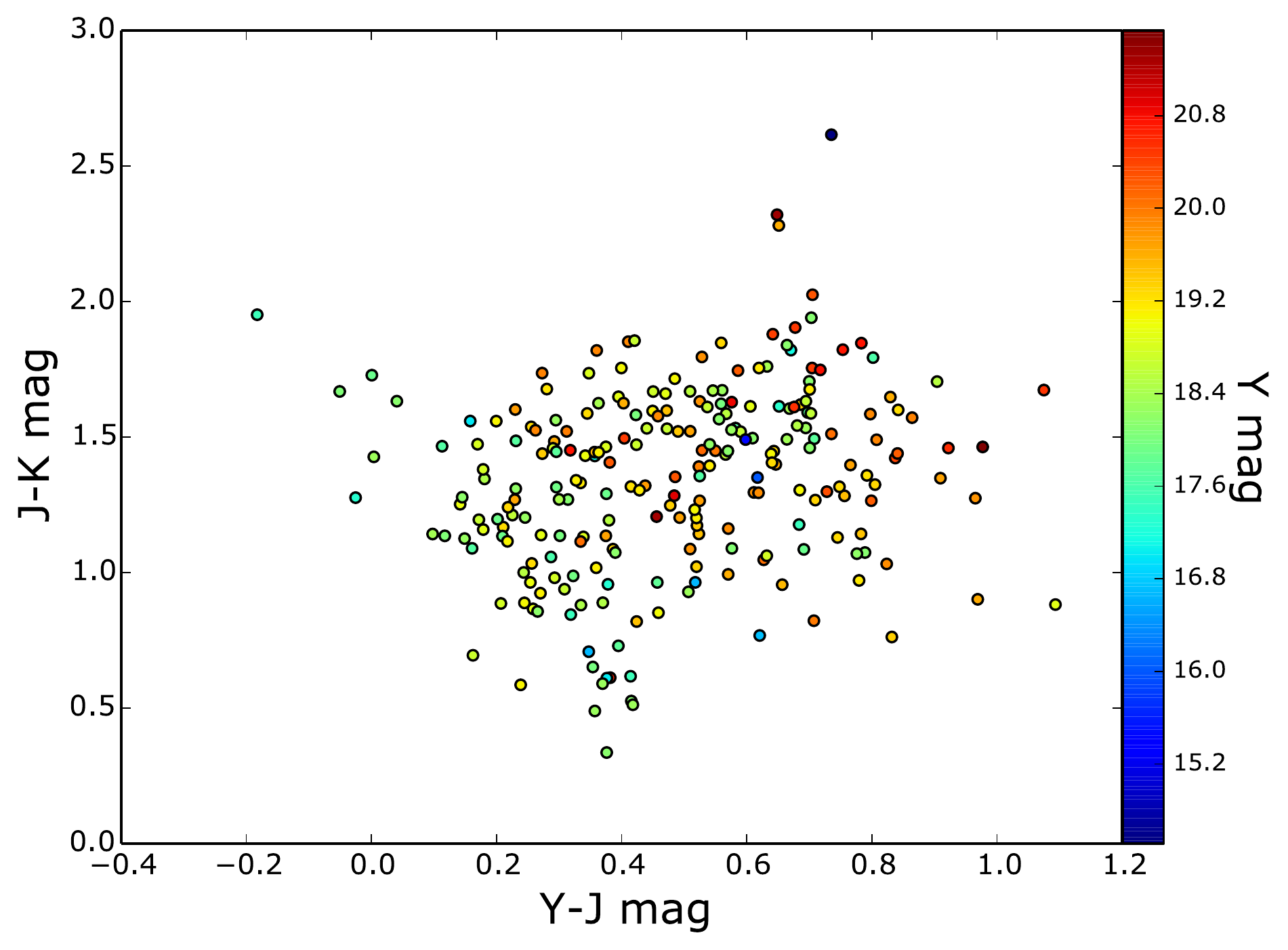}}


\caption{Distribution of the sample in the VISTA two-colour selection plane \citep{2013A&A...549A..29C}. The two figures are 
representative of the distribution of the sample in the VISTA plane, and the variation with  $J$ and $K_{\rm s}$ magnitudes show a similar pattern. 
X-ray flux ($F_{x}$) as in Fig.~\ref{allwise}.}
\label{vmc}
\end{figure*}

To investigate the distribution of the sources in the mid-infrared (ALLWISE) and near-infrared (VISTA) parameter space as a function of the  total X-ray flux (0.2--12 keV) and the infrared (ALLWISE and VISTA) magnitudes, we plotted colour--colour diagram in Figs.~\ref{allwise} and \ref{vmc}.
We found that the sample is homogeneously distributed within the ALLWISE ($x \equiv \log \frac{(f12\mu m)}{(f4.6 \mu m)}$ versus $y \equiv \log \frac{(f4.6\mu m)}{(f3.4 \mu m)}$)
and VISTA ($Y-J$ and $J-K$) colour--colour space as a function of their X-ray fluxes and the ALLWISE and VISTA magnitudes. 
The median of the ALLWISE magnitudes are 15.79, 14.88 and 11.88 mag in the W1, W2 and W3 bands and the corresponding standard deviations are 1.06, 1.00 and 0.92 mag respectively.
The corresponding median values of the VISTA magnitudes in the $Y$, $J$ and $K_{\rm s}$ bands are 19.00, 18.58 and 17.27 mag with standard deviations of 1.03, 1.1 and 1.06 mag respectively.

A positive correlation is observed between the integrated X-ray flux (0.2--12 keV) and
the ALLWISE and VISTA magnitudes. From Fig.~\ref{xray-ir} it can be seen that the brightest objects in X-rays correspond to the brightest objects in the W1, W2 and W3
bands of the ALLWISE data, and the $Y$, $J$, and $K_{\rm s}$ magnitudes in the VISTA data. 
\subsection{Identification of new candidates}
We identified 81 new AGN candidates from a sample that was selected using the
ALLWISE mid-infrared selection criteria of \cite{2012MNRAS.426.3271M} albeit with a lower S/N of 3, but with a corresponding
X-ray association. This sample is not included in any of the large all-sky AGN catalogues of \cite{2015ApJS..221...12S} and \cite{2015PASA...32...10F,2017yCat.7277....0F}. 
To ensure that they have not been reported previously, we searched within 3$\arcsec$ of all the 
source positions in the
VizieR database\footnote{http://vizier.u-strasbg.fr/}. We found that for 31 of the 81 sources, 
the corresponding
ALLWISE counterparts were identified
in the ARCHES (Astronomical Resource Cross-matching for High Energy Studies) database
\citep{2016arXiv160900809M} although source classification was not performed. 
22 of the 31 were however included in \cite{2009ApJ...701..508K} as AGN candidates. 
This makes 59 new sources catalogued as AGN candidates for the first time.
 The X-ray, mid-infrared and near-infrared colours of the sources, as well as the variability in the near-infrared band strongly support their AGN nature and the level of contamination
is likely to be almost negligible.
The new candidates therefore make a very promising subset for optical spectroscopic follow-up and are listed in
Table \ref{tab:new}.

\subsection{Optical spectroscopy of new candidates}
In order to verify the robustness of our selection of new candidates, we performed optical 
spectroscopic follow-up of 6 randomly chosen objects out of the 15 brightest (in X-rays) candidates
listed in Table \ref{tab:new}. 
Low-resolution (R=$\lambda$/$\Delta$$\lambda$$\sim$440) spectra over 
$\lambda$4450-8650$\AA$ were obtained with the FOcal Reducer and low 
dispersion Spectrograph \citep[FORS2;][]{1998Msngr..94....1A} at the 
ESO Very Large Telescope. The set up was: long-slit mode, grism 300V+10, 
order sorting filter GG435+81, slit width 1.3\arcsec. Further details 
are listed in Table\,\ref{tab:obslog}. The data reduction closely 
followed \citet{2016A&A...588A..93I}, except a newer Reflex-based 
\citep{2013A&A...559A..96F} ESO pipeline version (5.3.23) was used. 
The finding charts are shown in Fig.\,\ref{fig:fcs}, the final spectra are shown 
in Fig.\,\ref{fig:spectra}, and the measured emission lines are listed 
in Table\,\ref{tab:lines}. The candidates are presented in decreasing order of their X-ray fluxes. 
The results confirm that all of the 6 objects that were followed up are AGN. One of them (candidate 59) is a distant quasar
with $z=2.23$. The presence of a broad Mg{\sc ii} line in candidates 18, 26 and 35 also confirms that they are quasars. Candidates 06 and 47 are Seyfert galaxies, and the presence of narrow emission lines indicate type 2 Seyferts. 
Although the 100\% success rate of our initial follow-up of the 6 candidates does not guarantee the validity of all the other candidates,
the results are in further support of the robustness of our selection and strongly encourage future follow-ups.
\subsection{Comparison of the sample with the expected population of AGN}
\label{sec:comparison}
Seventy-one percent of the X-ray point sources detected in the SMC region by the \xmm\, are expected to be AGN behind the SMC
\citep{2013A&A...558A...3S}. This implies that $\sim$3158 of the 4311 detected unique sources are AGN. Our identification of 276 sources constitute only a 
small fraction of the entire expected population.
This can be attributed to the limitations of the selection criterion used in this work as well as the 
data sets which were used for the identification of the sample. 

The AGN selection wedge is defined using the Bright Ultra-hard \xmm\, survey (BUXS).
This is one of the largest complete flux-limited samples of bright (flux between 4.5--10 keV $>$ 6.0\ergcm{-14}) and ultra-hard (4.5--10 keV) X-ray selected
AGN. The sample selected in this work is heavily biased towards these properties which is evident by the fact that the median flux of the sample
is very close to the BUXS value. \cite{2012MNRAS.426.3271M} further noted that the selection completeness was significantly smaller for type 2 AGN 
at X-ray luminosities (2--10 keV
) $< 10^{44}$ erg s$^{-1}$ than type 1 AGN. This is mainly attributed to different luminosity distributions for the two classes of AGN, with type 2 being intrinsically
less luminous. Further a sharp decrease in the X-ray detection fraction of ALLWISE objects is expected 
at $\log \frac{(f12\mu m)}{(f4.6 \mu m)} \gtrsim 0.7-0.8$. This is also seen in our 
sample (Fig.~\ref{allwise-comp}), when
comparing the AGN identified using the ALLWISE criterion in this work with the entire sample of ALLWISE sources located
within the area of the \xmm\, survey which fulfilled the criterion of \cite{2012MNRAS.426.3271M}.

\begin{figure*}
\resizebox{0.33\hsize}{!}{\includegraphics[clip=]{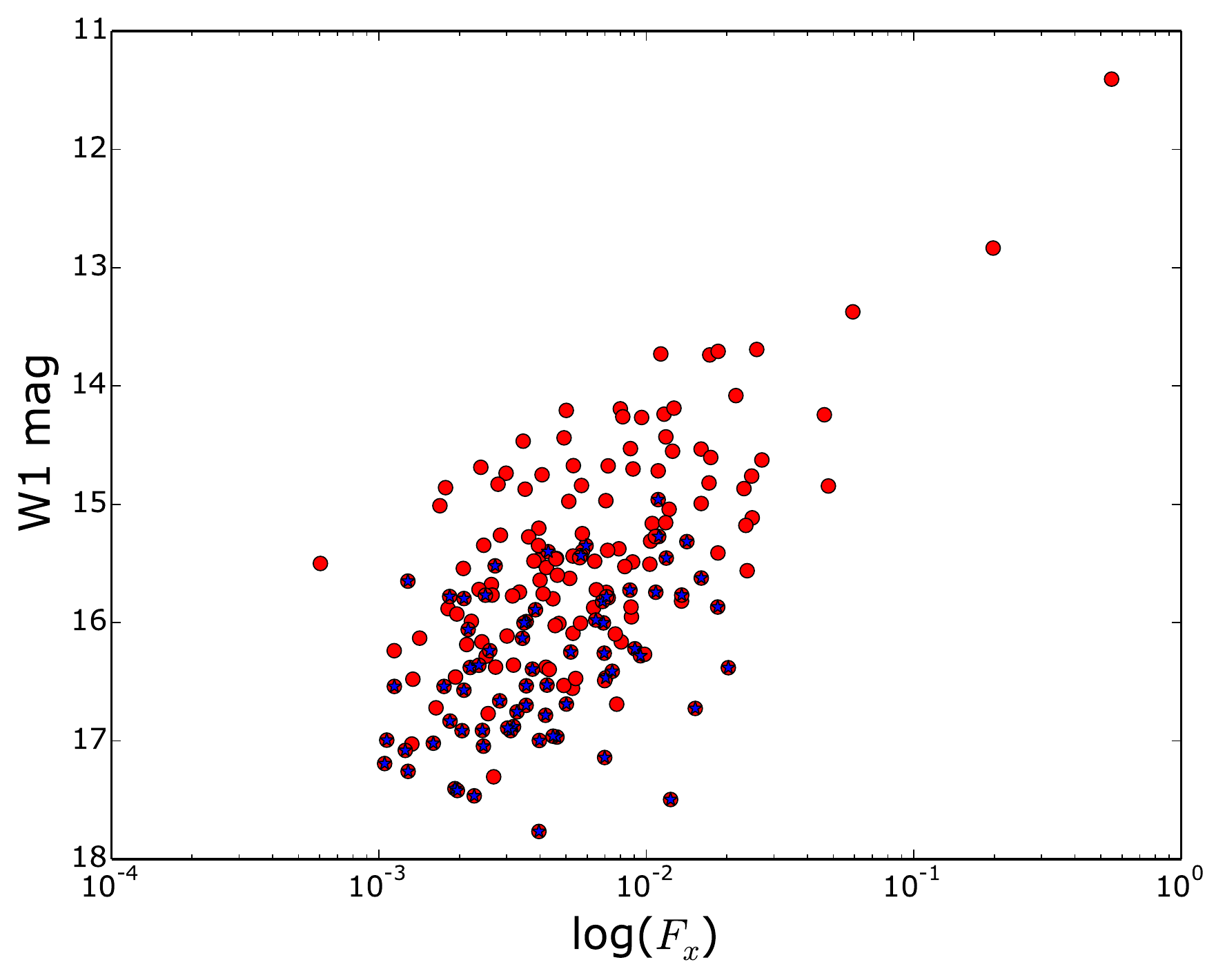}}
\resizebox{0.33\hsize}{!}{\includegraphics[clip=]{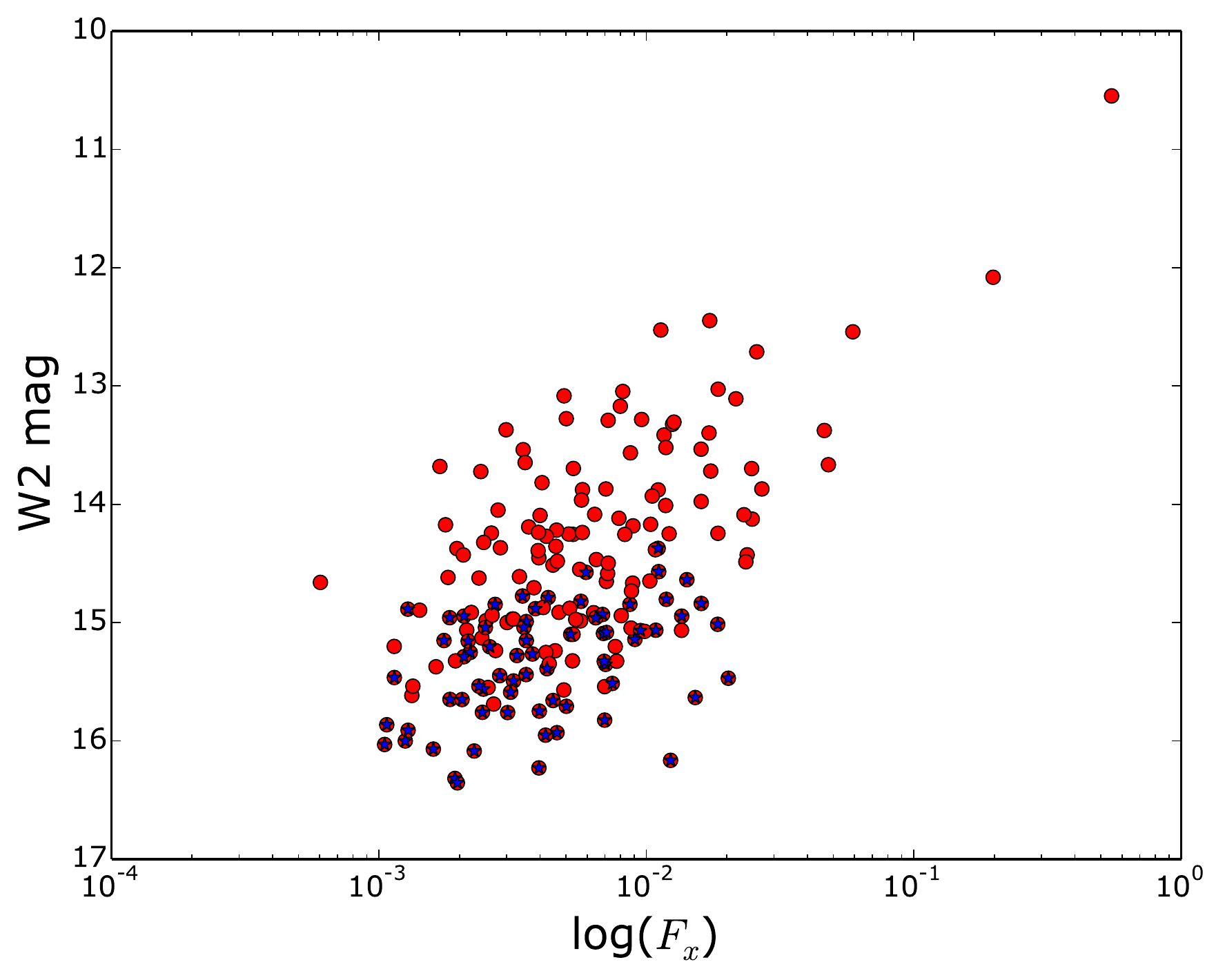}}
\resizebox{0.33\hsize}{!}{\includegraphics[clip=]{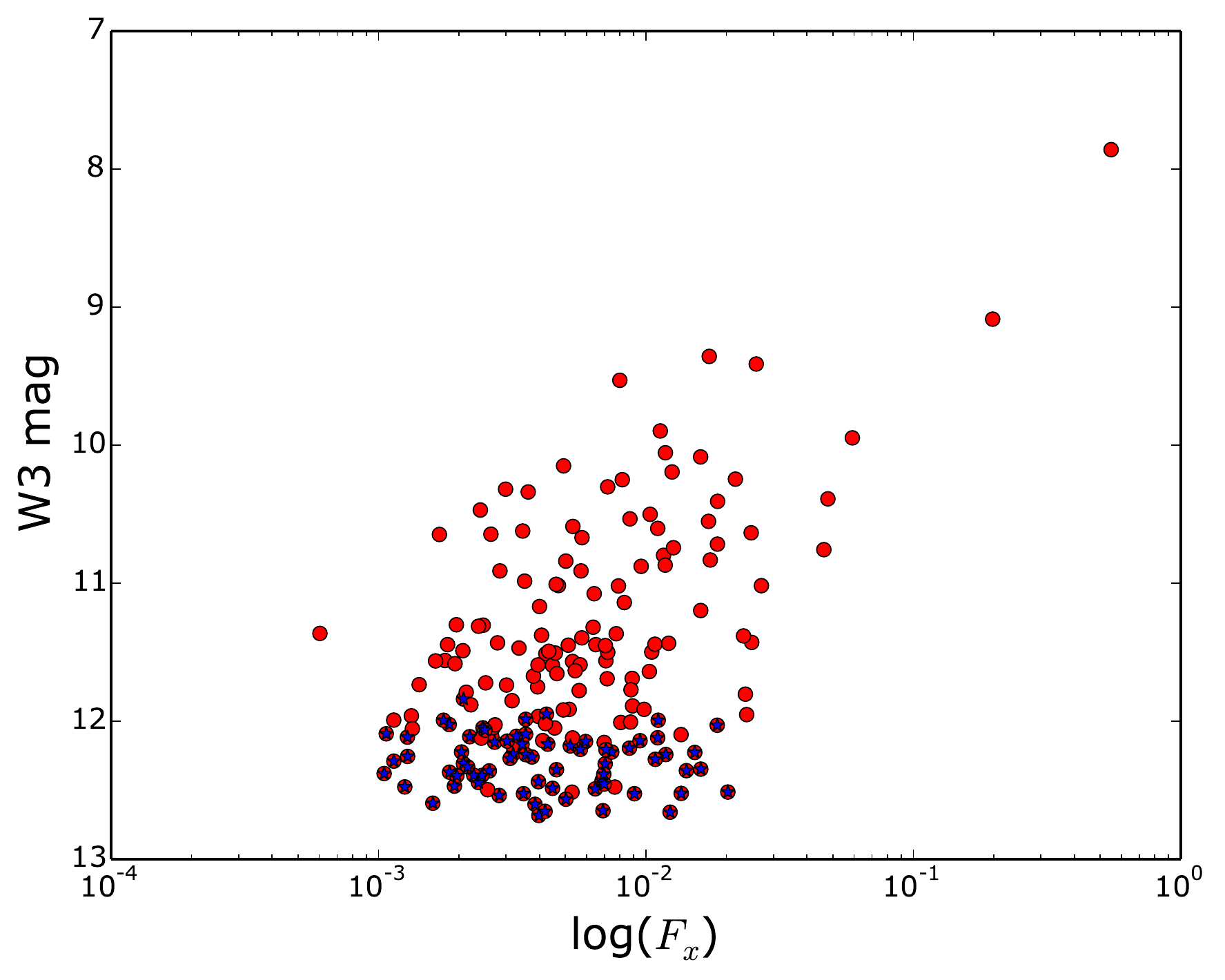}}

\resizebox{0.33\hsize}{!}{\includegraphics[clip=]{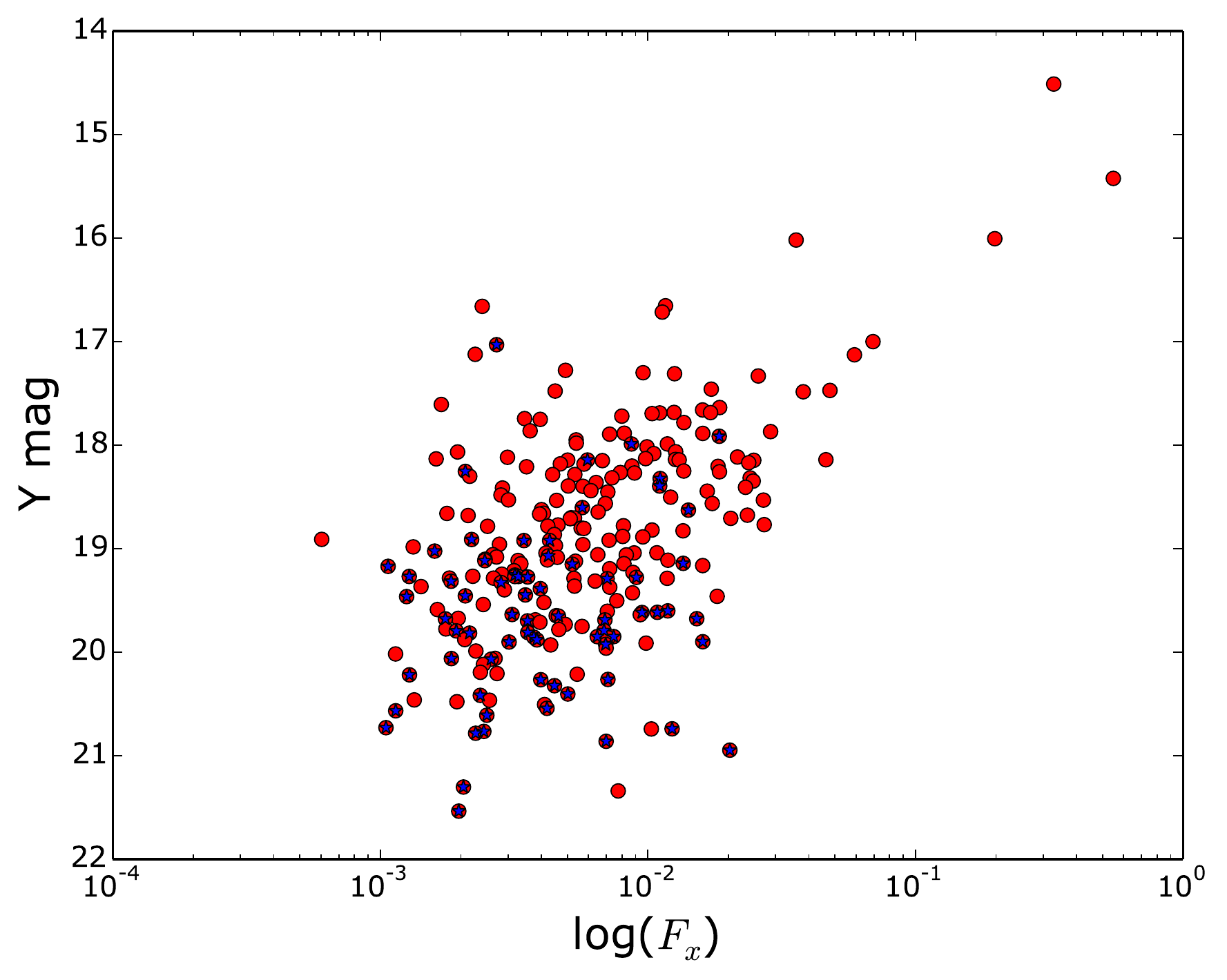}}
\resizebox{0.33\hsize}{!}{\includegraphics[clip=]{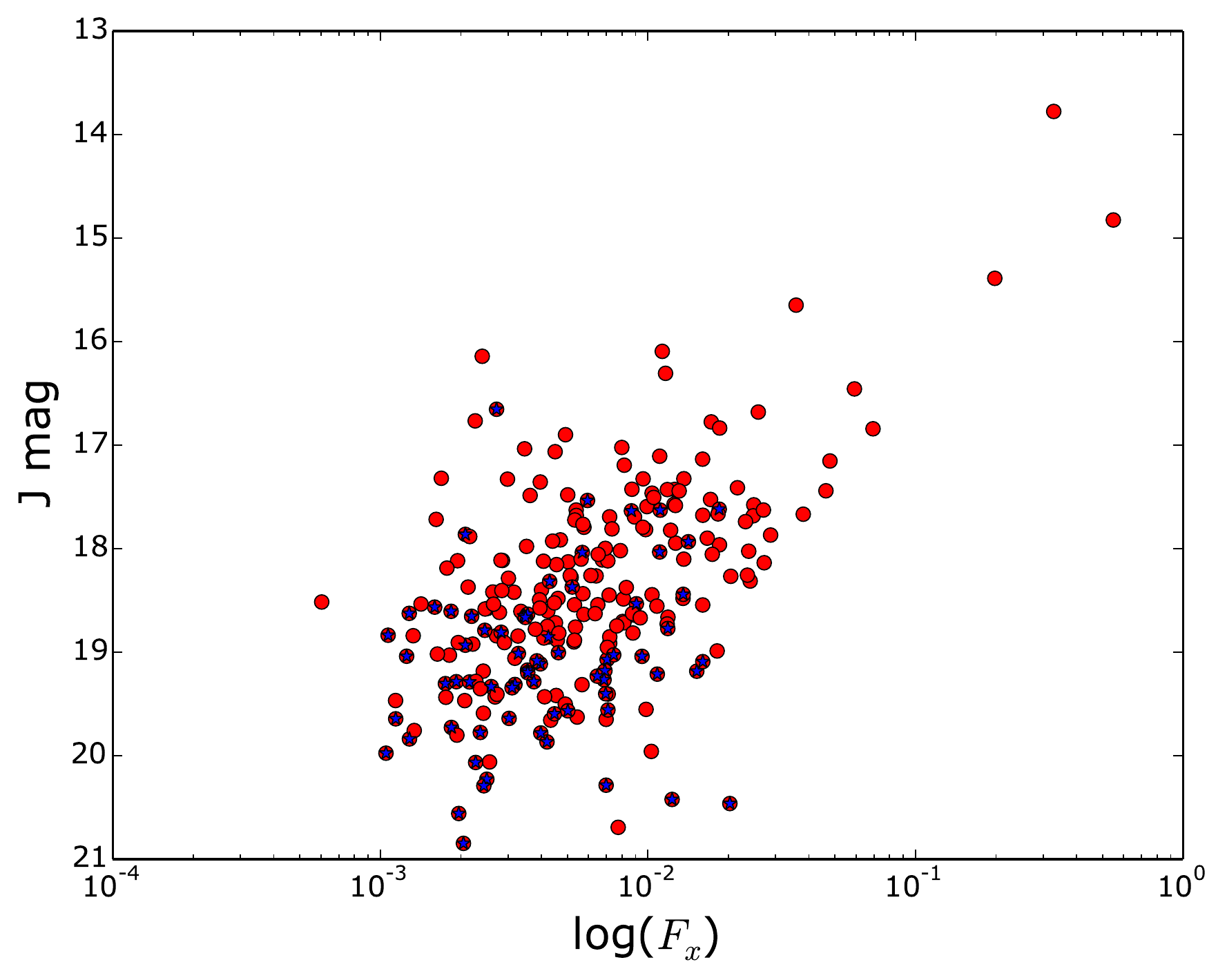}}
\resizebox{0.33\hsize}{!}{\includegraphics[clip=]{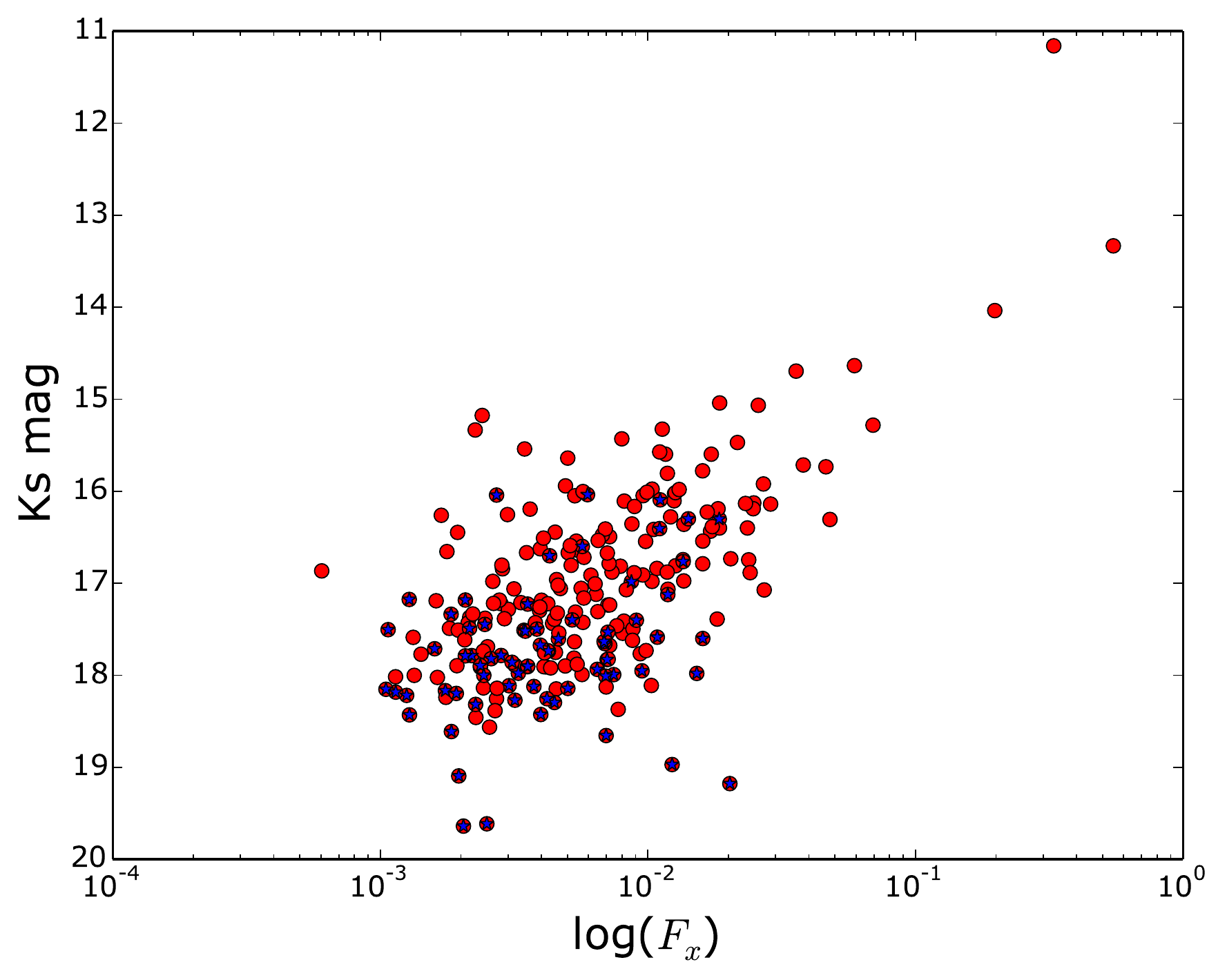}}
\caption{Relation between the X-ray flux and the ALLWISE and VISTA magnitudes.
         X-ray flux ($F_{x}$) as in Fig.~\ref{allwise}. Blue dots mark the new candidates identified in this work.}
\label{xray-ir}
\end{figure*}

\begin{figure*}
\centering
\resizebox{1.0\hsize}{!}{\includegraphics[]{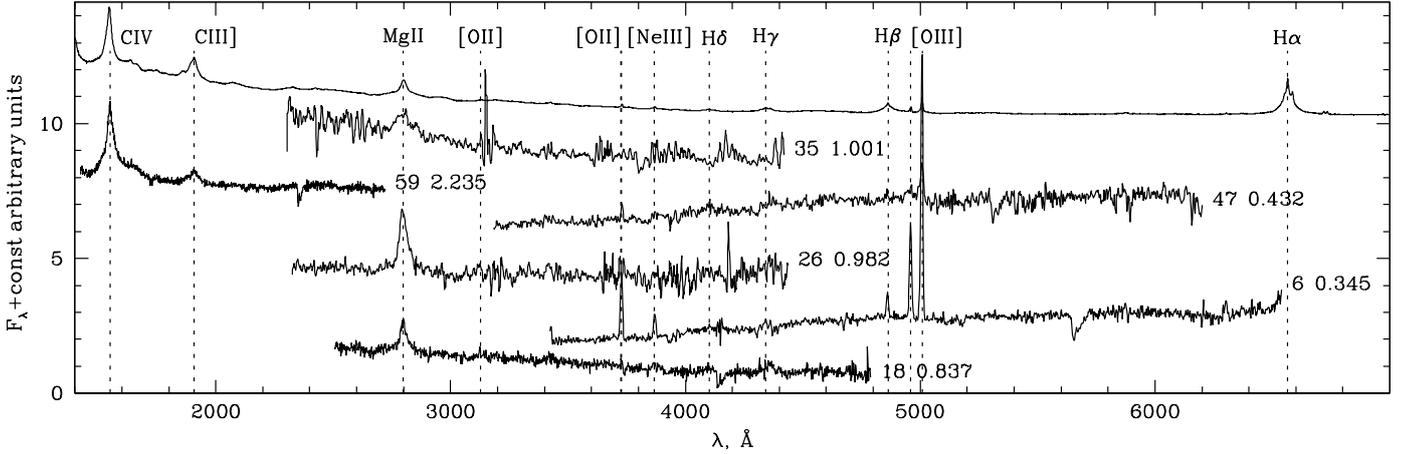}}
\caption{Spectra of the AGN candidates shifted to rest-frame wavelength. 
The spectra were smoothed, normalised to an average value of one and offset 
vertically for display purposes. The SDSS composite quasar spectrum 
\citep{2001AJ....122..549V} is shown at the top.}
\label{fig:spectra}
\end{figure*}

\begin{figure*}
\centering
\resizebox{0.5\hsize}{!}{\includegraphics[]{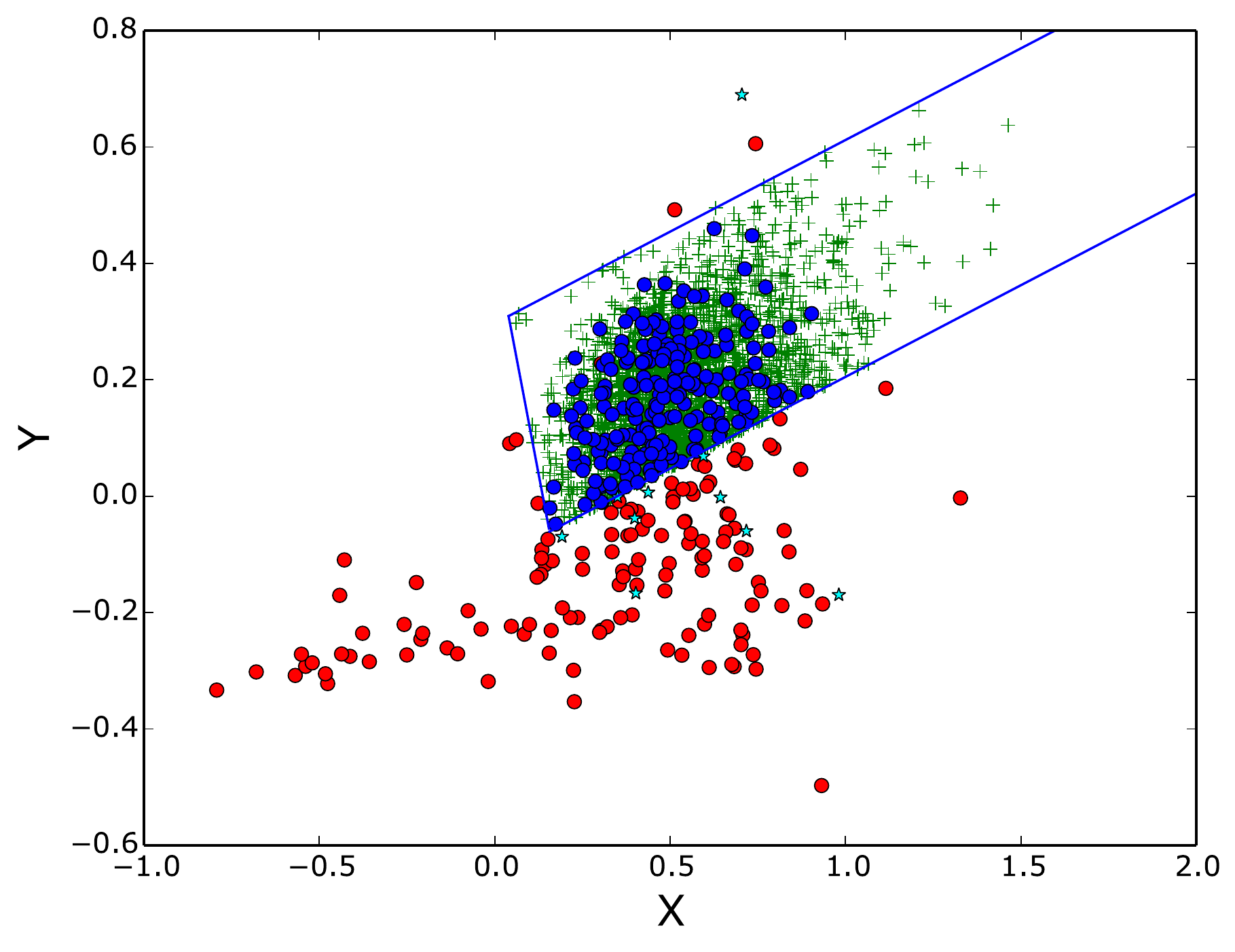}}
\caption{Distribution of sources in the ALLWISE two-colour selection plane \citep{2012MNRAS.426.3271M}.
$x \equiv \log \frac{(f12\mu m)}{(f4.6 \mu m)}$ and $y \equiv \log \frac{(f4.6\mu m)}{(f3.4 \mu m)}$.
Blue circles mark the 218 AGN identified using the ALLWISE criteria. Green crosses indicate all ALLWISE 
sources located within the \xmm\ survey area which fulfilled the criterion of \cite{2012MNRAS.426.3271M}.
Red circles are AGN candidates from \cite{2013A&A...558A...3S} which were found to have an ALLWISE counterpart.
Cyan stars are HMQ/MILLIQUAS sources which were found to have an ALLWISE counterpart.}
\label{allwise-comp}
\end{figure*}

In order to illustrate the limitations of the data sets which were used for the identification of the sample, we correlated the
sample of AGN candidates from \cite{2013A&A...558A...3S} which were not confirmed in this work, with the ALLWISE catalogue
(using sources with  $S/N \geq 3$ in W1, W2 and W3 bands \& cc$\_$flags ==`0000'). Out of 1989 sources only 134 were found to have a secure ALLWISE
counterpart (shown in Fig.\,\ref{allwise-comp}). The small fraction of the AGN candidates identified
with an ALLWISE counterpart is consistent with the fact that only 276 AGN were identified 
when correlating the entire unique source list of the \xmm\, SMC survey with the
ALLWISE catalogue. The most intriguing point is that almost all of these sources overlap with the horizontal sequence of normal galaxies in the
ALLWISE colour--colour space. This points to the fact that most of these 134 sources being at the faint end of the X-ray flux (near the detection limit
of the \xmm\, survey)  may have  
contribution from the host galaxy in the mid-infrared emission, so that they fall outside the selection
wedge and overlap with normal galaxies \citep[and references therein]{2012MNRAS.426.3271M,2017MNRAS.468.3042M}. 
This is further ascertained in Fig.~\ref{xray-ir}, which shows the new AGN candidates which are identified in this work by relaxing the S/N
threshold in the W1, W2 and W3 bands. It is evident from the figure that these sources tend to be intrinsically
less luminous, and they lie near the detection thresholds in both the X-ray and infrared bands. As the correlation between the X-ray and infrared
bands has a relatively large spread, it is possible that many of these sources fall below the detection threshold in the ALLWISE bands even though they are detected in the X-ray band. 
This is supported by the very small fraction of ALLWISE counterparts identified as AGN candidates from \cite{2013A&A...558A...3S}.
We further searched for the ALLWISE counterparts of the 58 sources exclusively selected from the HMQ and MILLIQUAS catalogue using the same criteria as above.
Only 10 sources were found to have a secure ALLWISE counterpart, plotted in Fig.~\ref{allwise-comp}. As in the case of AGN candidates 
from \cite{2013A&A...558A...3S},
all of these sources lie in the horizontal sequence of normal galaxies in the
ALLWISE colour--colour space. This is in further support of the fact that not all AGN lie within the ALLWISE selection wedge of \cite{2012MNRAS.426.3271M}.



%
%
\section{Summary \& Conclusions}
\label{sec:conclusion}

In this paper we present a catalogue of 276 AGN behind the SMC by correlating an updated X-ray point source catalogue of 
the \xmm\ SMC survey with already 
known AGN from the literature or selected using ALLWISE mid-infrared colour criteria. 
Ninety sources in the sample have confirmed redshifts, and the redshift distribution indicates that we detect sources from nearby Seyfert galaxies to distant quasars.
We investigated the X-ray luminosity distribution (not corrected for absorption) of the sources and compared it with the infrared luminosity distribution (both in the rest-frame). 

We identified near-infrared counterparts of the sources from the VISTA observations, and confirmed that the VISTA colours and variability are compatible with that expected from AGN.
We also confirm that the X-ray hardness ratios are compatible with that expected typically for AGN. These two results validate the robustness of our selection. 
We found the sources to be homogeneously distributed in the ALLWISE and VISTA colour--colour space
where the AGN are expected to be located.  A positive correlation was observed between the integrated X-ray flux (0.2--12 keV) and
the ALLWISE and VISTA magnitudes.

The sample contains 81 newly identified candidates (with 59 being introduced for the first time as AGN candidates) which were selected using the ALLWISE mid-infrared 
colours but with a lower S/N criterion than used by \cite{2015ApJS..221...12S}. An initial optical spectroscopic follow-up of 6 out of the 81 candidates confirmed that all six are AGN.
This strongly supports the robustness of the sample of candidate AGN and encourages further follow-up of the rest of the sample.
We also identified possible candidates for distant obscured quasars behind the SMC. All of these make a very promising subset for optical spectroscopic follow-up to confirm the nature of the sources,
 and/or X-ray spectral analysis to determine the degree and nature of obscuration towards the source.

In this work $\sim$ 75\% of the sources are identified using mid-infrared colour selection criteria. This technique is very useful in finding
the most luminous and obscured AGN. However, the sample is expected to be incomplete at lower luminosities where the \bibliography{agn}

contamination from the host galaxy begins to dominate and renders the mid-infrared colours bluer so that they fall outside the selection
wedge \citep[and references therein]{2012MNRAS.426.3271M,2017MNRAS.468.3042M}. Therefore, we possibly miss the faint AGN using these criteria which become indistinguishable in X-rays from star-forming galaxies.

\begin{acknowledgements}
The \xmm\ project is supported by the Bundesministerium f{\"u}r Wirtschaft und 
Technologie/Deutsches Zentrum f{\"u}r Luft- und Raumfahrt (BMWI/DLR, FKZ 50 OX 0001)
and the Max-Planck Society. M.-R.L.C. has received support from the European Research Council (ERC) under European Union’s
Horizon 2020 research and innovation programme (grant agreement No 682115). We thank the Cambridge Astronomy Survey Unit (CASU) 
and the Wide Field Astronomy Unit (WFAU) in Edinburgh for providing calibrated data products under the support of the Science 
and Technology Facility Council (STFC).
\end{acknowledgements}

\bibliographystyle{aa}
\bibliography{agn}

\begin{appendix}

\section{Tables}


%

\onecolumn

\begin{landscape}
{\tiny
\begin{longtable}{cccccccccccc}
\caption{Newly identified AGN candidates from the \xmm\ SMC survey selected using ALLWISE mid-infrared colour selection.}
\label{tab:new}
\\
\hline\hline
Candidate & Designation & R.A. & Dec. & Error & X & Y & F$_{x}$ (0.2--12 keV) & eF$_{x}$ & $Y-J$ & $J-K_{\rm s}$ & {\it XMM}\_id \\
-- & $\circ$ & $\circ$ & $\arcsec$ & $\frac{(f12\mu m)}{(f4.6 \mu m)}$ & $\log \frac{(f4.6\mu m)}{(f3.4 \mu m)}$ &  \ergcm{-11} & erg cm$^{-2}$ s$^{-1}$ & mag & mag & --\\
\hline
\endfirsthead

\multicolumn{11}{c}%
{{\bfseries \tablename\ \thetable{} -- continued from previous page}} \\
\hline
Candidate & Designation & R.A. & Dec. & Error & X & Y & F$_{x}$ (0.2--12 keV) & eF$_{x}$ & $Y-J$ & $J-K_{\rm s}$ & {\it XMM}\_id \\
--          & $\circ$ & $\circ$ & $\arcsec$ & $\log \frac{(f12 \mu m)}{(f4.6 \mu m)}$ & $\log \frac{(f4.6 \mu m)}{(f3.4 \mu m)}$ & \ergcm{-11} & erg cm$^{-2}$ s$^{-1}$ & mag & mag & --\\
\hline
\endhead

\hline \multicolumn{12}{l}{{\tiny Continued on next page}} \\ 
\endfoot

\hline 
\endlastfoot
1 & J002523.92$-$712322.4	&	6.3497	&	$-$71.3896	&	0.1059	&	0.244	&	0.152	&	0.0065	&	0.0010	&	0.619	&	1.290	&	-	\\
2 & J002606.42$-$711301.6	&	6.5267	&	$-$71.2171	&	0.1505	&	0.633	&	0.200	&	0.0042	&	0.0007	&	0.217	&	1.120	&	-	\\
3 & J002636.61$-$712409.5	&	6.6525	&	$-$71.4027	&	0.1365	&	0.513	&	0.195	&	0.0022	&	0.0002	&	0.258	&	0.866	&	-	\\
4 & J002658.65$-$712104.2	&	6.7444	&	$-$71.3512	&	0.2273	&	0.604	&	0.271	&	0.0070	&	0.0004	&	0.576	&	1.630	&	-	\\
5 & J003816.75$-$725101.0	&	9.5698	&	$-$72.8503	&	0.1011	&	0.430	&	0.073	&	0.0018	&	0.0004	&	0.710	&	1.270	&	J003815.9$-$725058	\\
6 & J003918.19$-$730331.5	&	9.8258	&	$-$73.0588	&	0.0880	&	0.168	&	0.016	&	0.0140	&	0.0020	&	0.695	&	1.630	&	J003917.7$-$730330	\\
7 & J004140.61$-$735156.8	&	10.4192	&	$-$73.8658	&	0.0867	&	0.305	&       $-$0.011	&	0.0043	&	0.0009	&	0.606	&	1.610	&	-	\\
8 & J004551.17$-$734018.6	&	11.4632	&	$-$73.6719	&	0.0886	&	0.303	&       $-$0.009	&	0.0050	&	0.0004	&	0.566	&	1.440	&	J004551.3$-$734017	\\
9 & J004613.07$-$722539.5	&	11.5545	&	$-$72.4276	&	0.1894	&	0.619	&	0.181	&	0.0150	&	0.0020	&	0.493	&	1.200	&	J004612.9$-$722535	\\
10 & J004633.42$-$734443.2	&	11.6393	&	$-$73.7453	&	0.1022	&	0.334	&	0.014	&	0.0027	&	0.0005	&	0.376	&	0.611	&	J004632.9$-$734444	\\
11 & J004825.64$-$720032.6	&	12.1069	&	$-$72.0091	&	0.2030	&	0.520	&	0.285	&	0.0044	&	0.0006	&	0.0	&	1.790	&	J004825.8$-$720032	\\
12 & J004914.34$-$744537.6	&	12.3098	&	$-$74.7605	&	0.0849	&	0.317	&	0.096	&	0.0087	&	0.0010	&	0.354	&	0.652	&	$-$	\\
13 & J004929.63$-$721051.5	&	12.3735	&	$-$72.1810	&	0.1274	&	0.425	&	0.204	&	0.0052	&	0.0005	&	0.780	&	0.971	&	J004929.6$-$721050	\\
14 & J004938.79$-$715052.5	&	12.4116	&	$-$71.8479	&	0.2460	&	0.717	&	0.209	&	0.0010	&	0.0002	&	0.754	&	1.820	&	J004938.4$-$715052	\\
15 & J005015.69$-$743801.6	&	12.5654	&	$-$74.6338	&	0.0978	&	0.300	&	0.287	&	0.0034	&	0.0004	&	0.271	&	1.140	&	J005015.4$-$743802	\\
16 & J005019.50$-$714817.2	&	12.5813	&	$-$71.8048	&	0.1070	&	0.387	&	0.105	&	0.0022	&	0.0005	&	0.528	&	1.800	&	J005019.0$-$714818	\\
17 & J005035.42$-$730815.7	&	12.6476	&	$-$73.1377	&	0.4217	&	0.717	&	0.140	&	0.0007	&	0.0001	&	0.0	&	0.0	&	J005035.6$-$730813	\\
18 & J005057.59$-$714132.1	&	12.7400	&	$-$71.6923	&	0.1348	&	0.573	&	0.103	&	0.0075	&	0.0010	&	0.824	&	1.030	&	J005057.3$-$714131	\\
19 & J005100.96$-$735011.9	&	12.7540	&	$-$73.8366	&	0.5307	&	0.903	&	0.314	&	0.0016	&	0.0005	&	0.0	&	2.360	&	J005101.2$-$735011	\\
20 & J005141.19$-$750049.6	&	12.9216	&	$-$75.0138	&	0.1727	&	0.584	&	0.274	&	0.0031	&	0.0005	&	0.292	&	1.480	&	J005141.0$-$750051	\\
21 & J005145.45$-$720911.1	&	12.9394	&	$-$72.1531	&	0.1997	&	0.559	&	0.299	&	0.0032	&	0.0003	&       $-$0.050	&	1.040	&	J005145.3$-$720910	\\
22 & J005153.04$-$715334.9	&	12.9710	&	$-$71.8930	&	0.2001	&	0.627	&	0.250	&	0.0020	&	0.0005	&	0.456	&	1.210	&	J005152.5$-$715340	\\
23 & J005202.89$-$720505.0	&	13.0121	&	$-$72.0847	&	0.3675	&	0.773	&	0.359	&	0.0040	&	0.0003	&	0.273	&	1.440	&	J005203.0$-$720505	\\
24 & J005240.23$-$745251.0	&	13.1677	&	$-$74.8809	&	0.1913	&	0.765	&	0.196	&	0.0011	&	0.0002	&	0.334	&	1.330	&	J005240.0$-$745252	\\
25 & J005301.41$-$713107.7	&	13.2559	&	$-$71.5188	&	0.1710	&	0.569	&	0.218	&	0.0018	&	0.0003	&	0.335	&	1.110	&	J005301.1$-$713105	\\
26 & J005326.06$-$714821.2	&	13.3586	&	$-$71.8059	&	0.0853	&	0.281	&	0.005	&	0.0120	&	0.0008	&	0.830	&	1.650	&	J005325.9$-$714821	\\
27 & J005410.60$-$705051.0	&	13.5442	&	$-$70.8475	&	0.2896	&	0.659	&	0.277	&	0.0120	&	0.0030	&	0.318	&	1.450	&	J005411.6$-$705052	\\
28 & J005415.32$-$743153.0	&	13.5638	&	$-$74.5314	&	0.1259	&	0.459	&	0.196	&	0.0037	&	0.0003	&	0.570	&	1.160	&	J005415.5$-$743153	\\
29 & J005456.84$-$720857.9	&	13.7368	&	$-$72.1494	&	0.1349	&	0.312	&	0.154	&	0.0062	&	0.0007	&	0	&	0	&	J005456.6$-$720856	\\
30 & J005540.46$-$720555.2	&	13.9186	&	$-$72.0987	&	0.1624	&	0.449	&	0.258	&	0.0021	&	0.0002	&	0.524	&	1.140	&	J005540.4$-$720555	\\
31 & J005611.06$-$703958.6	&	14.0461	&	$-$70.6663	&	0.3069	&	0.841	&	0.170	&	0.0020	&	0.0005	&	0.977	&	1.460	&	J005610.7$-$703957	\\
32 & J005622.87$-$743238.6	&	14.0953	&	$-$74.5441	&	0.1209	&	0.434	&	0.117	&	0.0070	&	0.0009	&	0.523	&	1.390	&	J005623.7$-$743236	\\
33 & J005652.51$-$721203.5	&	14.2188	&	$-$72.2010	&	0.2488	&	0.688	&	0.159	&	0.0046	&	0.0002	&	0.647	&	1.400	&	J005652.4$-$721203	\\
34 & J005652.57$-$712300.6	&	14.2191	&	$-$71.3835	&	0.1787	&	0.525	&	0.265	&	0.0045	&	0.0005	&	0.728	&	1.300	&	J005652.3$-$712300	\\
35 & J005749.75$-$711802.3	&	14.4573	&	$-$71.3006	&	0.0819	&	0.253	&	0.058	&	0.0160	&	0.0007	&	0.808	&	1.490	&	J005749.3$-$711801	\\
36 & J005811.52$-$714748.4	&	14.5480	&	$-$71.7968	&	0.1634	&	0.595	&	0.248	&	0.0036	&	0.0005	&	0.525	&	1.270	&	J005811.3$-$714748	\\
37 & J010004.54$-$711738.8	&	15.0190	&	$-$71.2941	&	0.1313	&	0.439	&	0.109	&	0.0200	&	0.0020	&	0.484	&	1.280	&	J010004.2$-$711737	\\
38 & J010033.38$-$715037.2	&	15.1391	&	$-$71.8437	&	0.1435	&	0.494	&	0.074	&	0.0024	&	0.0005	&	0.642	&	1.880	&	J010033.4$-$715039	\\
39 & J010035.78$-$714702.8	&	15.1491	&	$-$71.7841	&	0.2359	&	0.647	&	0.124	&	0.0016	&	0.0003	&	0.459	&	0.852	&	J010036.1$-$714700	\\
40 & J010053.72$-$711042.9	&	15.2239	&	$-$71.1786	&	0.1030	&	0.459	&	0.144	&	0.0036	&	0.0005	&	0.640	&	1.410	&	J010053.0$-$711042	\\
41 & J010109.33$-$723237.0	&	15.2889	&	$-$72.5436	&	0.0820	&	0.287	&	0.026	&	0.0110	&	0.0006	&	0.694	&	1.530	&	J010109.3$-$723237	\\
42 & J010140.41$-$722940.0	&	15.4184	&	$-$72.4945	&	0.1103	&	0.386	&	0.079	&	0.0028	&	0.0004	&	0.376	&	0.336	&	J010140.6$-$722941	\\	
43 & J010155.58$-$722948.4	&	15.4816	&	$-$72.4968	&	0.0911	&	0.228	&	0.054	&	0.0059	&	0.0006	&	0.609	&	1.500	&	J010155.5$-$722948	\\
44 & J010231.92$-$730709.3	&	15.6330	&	$-$73.1193	&	0.1179	&	0.424	&	0.057	&	0.0081	&	0.0004	&	0.805	&	1.325	&	J010231.8$-$730708	\\
45 & J010525.62$-$721929.7	&	16.3568	&	$-$72.3249	&	0.1545	&	0.475	&	0.054	&	0.0031	&	0.0003	&	0.0	&	1.640	&	J010525.4$-$721927	\\
46 & J010630.55$-$723544.0	&	16.6273	&	$-$72.5956	&	0.1010	&	0.499	&	0.085	&	0.0021	&	0.0002	&	0.390	&	0.680	&	J010630.7$-$723546	\\
47 & J010849.53$-$721233.1	&	17.2064	&	$-$72.2092	&	0.0949	&	0.225	&	0.073	&	0.0140	&	0.0004	&	0.701	&	1.680	&	J010849.7$-$721233	\\
48 & J010927.41$-$723110.0	&	17.3642	&	$-$72.5195	&	0.1186	&	0.362	&	0.266	&	0.0050	&	0.0005	&	0.0	&	2.130	&	J010927.7$-$723109	\\
49 & J010934.35$-$720338.0	&	17.3931	&	$-$72.0606	&	0.1029	&	0.355	&	0.018   &	0.019	&	0.0006	&	0.864	&	1.571	&	J010934.5$-$720337	\\	
50 & J011014.66$-$724228.6	&	17.5611	&	$-$72.7080	&	0.2437	&	0.719	&	0.283	&	0.0013	&	0.0003	&	0.381	&	1.410	&	J011014.3$-$724228	\\
51 & J011241.93$-$732830.6	&	18.1747	&	$-$73.4752	&	0.1766	&	0.513	&	0.137	&	0.0050	&	0.0005	&	0.837	&	1.420	&	J011242.2$-$732831	\\
52 & J011323.84$-$732401.6	&	18.3494	&	$-$73.4005	&	0.1453	&	0.527	&	0.174	&	0.0011	&	0.0002	&	0.922	&	1.460	&	J011324.4$-$732400	\\
53 & J011335.15$-$735510.3	&	18.3965	&	$-$73.9195	&	0.1505	&	0.524	&	0.335	&	0.0033	&	0.0005	&	0.256	&	1.030	&	-	\\
54 & J011420.98$-$722328.4	&	18.5874	&	$-$72.3912	&	0.1812	&	0.703	&	0.197	&	0.0030	&	0.0004	&	0.262	&	1.530	&	J011421.3$-$722327	\\
54 & J011452.86$-$735439.2	&	18.7203	&	$-$73.9109	&	0.1004	&	0.263	&	0.129	&	0.0035	&	0.0003	&	0.783	&	1.140	&	-	\\
56 & J011543.81$-$735002.0	&	18.9326	&	$-$73.8339	&	0.1318	&	0.421	&	0.297	&	0.0036	&	0.0003	&	0.611	&	1.300	&	-	\\
57 & J011606.88$-$731725.0	&	19.0287	&	$-$73.2903	&	0.0981	&	0.169	&	0.148	&	0.0038	&	0.0009	&	0.798	&	1.580	&	J011606.7$-$731725	\\
58 & J011732.79$-$735228.6	&	19.3867	&	$-$73.8746	&	0.1794	&	0.576	&	0.077	&	0.0042	&	0.0020	&	0.676	&	1.610	&	-	\\
59 & J011744.71$-$733922.1	&	19.4363	&	$-$73.6562	&	0.0977	&	0.451	&	0.086	&	0.0180	&	0.0006	&	0.296	&	1.320	&	J011744.7$-$733923	\\
60 & J011915.12$-$733052.9	&	19.8130	&	$-$73.5147	&	0.1280	&	0.520	&	0.300	&	0.0018	&	0.0002	&	0.375	&	1.140	&	J011915.0$-$733054	\\
61 & J012007.47$-$734917.6	&	20.0311	&	$-$73.8216	&	0.0905	&	0.447	&	0.035	&	0.0025	&	0.0003	&	0.382	&	0.612	&	-	\\
62 & J012008.63$-$733045.0	&	20.0360	&	$-$73.5125	&	0.1037	&	0.234	&	0.109	&	0.0069	&	0.0004	&	0.510	&	1.520	&	J012008.8$-$733044	\\
63 & J012033.96$-$752359.1	&	20.1415	&	$-$75.3998	&	0.1064	&	0.427	&	0.230	&	0.0095	&	0.0010	&	0.580	&	1.080	&	-	\\
64 & J012142.29$-$732051.0	&	20.4262	&	$-$73.3475	&	0.1918	&	0.603	&	0.206	&	0.0024	&	0.0004	&	0.470	&	2.280	&	-	\\
65 & J012149.23$-$724949.3	&	20.4551	&	$-$72.8304	&	0.0920	&	0.257	&	0.100	&	0.0069	&	0.0007	&	0.525	&	1.630	&	-	\\
66 & J012149.94$-$724445.2	&	20.4581	&	$-$72.7459	&	0.0940	&	0.372	&	0.016	&	0.0110	&	0.0010	&	0.403	&	1.630	&	-	\\
67 & J012206.29$-$730450.1	&	20.5262	&	$-$73.0806	&	0.0861	&	0.365	&	0.049	&	0.0013	&	0.0002	&	0.643	&	1.450	&	J012205.8$-$730447	\\
68 & J012455.51$-$733001.9	&	21.2313	&	$-$73.5005	&	0.2014	&	0.483	&	0.244	&	0.0040	&	0.0002	&	0.486	&	1.350	&	-	\\
69 & J012806.97$-$730824.4	&	22.0291	&	$-$73.1401	&	0.1832	&	0.743	&	0.228	&	0.0015	&	0.0002	&	0	&	0.0	&	-	\\
70 & J012820.97$-$732203.4	&	22.0874	&	$-$73.3676	&	0.1870	&	0.662	&	0.337	&	0.0025	&	0.0003	&	0.327	&	1.340	&	-	\\
71 & J012938.05$-$742144.8	&	22.4086	&	$-$74.3625	&	0.1208	&	0.474	&	0.190	&	0.0070	&	0.0010	&	0.218	&	1.240	&	-	\\
72 & J013028.52$-$730004.9	&	22.6189	&	$-$73.0014	&	0.1673	&	0.520	&	0.170	&	0.0013	&	0.0003	&	0	&	2.070	&	-	\\
73 & J013119.92$-$742505.7	&	22.8330	&	$-$74.4183	&	0.1324	&	0.421	&	0.230	&	0.0028	&	0.0005	&	0.520	&	1.020	&	-	\\
74 & J013206.20$-$733254.7	&	23.0258	&	$-$73.5485	&	0.1131	&	0.394	&	0.158	&	0.0026	&	0.0001	&	0.735	&	1.510	&	-	\\
75 & J013209.11$-$743504.9	&	23.0380	&	$-$74.5847	&	0.0619	&	0.157	&       $-$0.020 	&	0.0110	&	0.0006	&	0.363	&	1.620	&	-	\\
76 & J013211.69$-$733621.0	&	23.0487	&	$-$73.6058	&	0.2683	&	0.734	&	0.296	&	0.0023	&	0.0003	&	0.718	&	1.750	&	-	\\
77 & J013254.61$-$743954.3	&	23.2276	&	$-$74.6651	&	0.0870	&	0.407	&	0.024	&	0.0071	&	0.0020	&	0.705	&	2.030	&	-	\\
78 & J013310.11$-$731129.9	&	23.2922	&	$-$73.1916	&	0.2723	&	0.795	&	0.179	&	0.0019	&	0.0003	&	0.510	&	1.090	&	-	\\
79 & J013349.39$-$731942.8	&	23.4558	&	$-$73.3286	&	0.2060	&	0.666	&	0.176	&	0.0013	&	0.0001	&	0.424	&	0.820	&	-	\\
80 & J013445.95$-$732005.7	&	23.6915	&	$-$73.3349	&	0.1151	&	0.303	&	0.176	&	0.0091	&	0.0005	&	0.745	&	1.130	&	-	\\
81 & J013607.46$-$741731.7	&	24.0311	&	$-$74.2922	&	0.1111	&	0.405	&	0.150	&	0.0120	&	0.0030	&	0.0	&	1.890	&	-	\\
\hline\hline
\end{longtable}
}

 \begin{center}

  { \scriptsize The designation, RA, Dec, associated error and the X and Y colours correspond to the ALLWISE source.}\\
  
  {\scriptsize The X-ray flux and the associated error (F$_{x}$ and eF$_{x}$) are from the corresponding X-ray counterpart.}\\
  
   {\scriptsize $Y-J$ and $J-K_{\rm s}$ colours are from the corresponding VISTA counterpart.}\\

  {\scriptsize Sources where $Y-J$ and $J-K_{\rm s}$ are 0 do not have a VISTA counterpart in the corresponding bands to compute the colours.}\\
  
  {\scriptsize {\it XMM}\_ids correspond to sources already included in \cite{2013A&A...558A...3S}.}

\end{landscape}

\twocolumn
\begin{table}
\caption[]{Log of the spectroscopic observations.
For each object we list: name, coordinates, UT at the start of the 
observations, number and exposure times for individual spectra, 
airmass range, slit's position angle and adopted redshift.}
\begin{center}
\begin{tabular}{@{ }lcccccc@{ }}
\hline\hline\noalign{\smallskip}
Candidate         & RA Dec                     & UT at start of obs. & Exp.         & sec\,$z$     & Slit Pos. & Adopted         \\
             & (J2000)                    & yyyy-mm-ddThh:mm:ss & (s)          & (dex)        & Ang. (deg)& Redshift        \\
\noalign{\smallskip}\hline\noalign{\smallskip}
59 & 01:17:44.713 $-$73:39:22.173 & 2017-10-22T00:03:49 & 2$\times$090 & 1.866--1.854 &    69.182 & 2.235$\pm$0.008 \\
35 & 00:57:49.754 $-$71:18:02.337 & 2017-10-05T00:18:44 & 1$\times$300 & 1.924--1.902 &    77.926 & 1.001$\pm$0.015 \\
06 & 00:39:18.199 $-$73:03:31.528 & 2017-10-23T05:30:37 & 2$\times$180 & 1.612--1.625 &    $-$37.557 & 0.345$\pm$0.001 \\
47 & 01:08:49.536 $-$72:12:33.138 & 2017-10-15T00:13:28 & 2$\times$210 & 1.870--1.843 &    74.465 & 0.432$\pm$0.002 \\
26 & 00:53:26.067 $-$71:48:21.210 & 2017-10-22T00:25:36 & 2$\times$270 & 1.690--1.666 &    59.565 & 0.982$\pm$0.015 \\
18 & 00:50:57.598 $-$71:41:32.166 & 2017-10-23T05:45:16 & 2$\times$300 & 1.582--1.603 &    $-$40.381 & 0.837$\pm$0.015 \\
\noalign{\smallskip}\hline
\end{tabular}
\end{center}
\label{tab:obslog}
\end{table}
\begin{table}
\caption[]{Derived line parameters from our optical spectra.
For the doublet [O{\sc ii}] that is unresolved in our data we 
adopted a rest wavelength of 3727.89\,\AA\ averaging the central 
wavelengths of the two components in 1:2 ratio, approximately 
proportional to the flux ratio of the two components. The 
wavelength and redshift errors reflect only random uncertainties.}
\begin{center}
\begin{tabular}{@{ }l@{ }c@{ }c@{ }c@{ }}
\hline\hline\noalign{\smallskip}
Candidate         & Spectral features &  observed wavelength ($\AA$)  & Redshift \\
\noalign{\smallskip}\hline\noalign{\smallskip}
59 & C{\sc iv}  &   5015.26$\pm$1.26 & 2.238$\pm$0.001 \\
               & C{\sc iii}  & 6164.46$\pm$4.63 & 2.230$\pm$0.002 \\
35 & Mg{\sc ii} &   5600.17$\pm$5.47 & 1.001$\pm$0.002 \\
06 & [O{\sc ii}] &  5013.71$\pm$0.05 & 0.345$\pm$0.001 \\
               & [Ne{\sc iii}] & 5204.39$\pm$0.05 & 0.345$\pm$0.001 \\
               & H$\beta$   &   6537.28$\pm$0.03 & 0.344$\pm$0.001 \\
               & [O{\sc iii}] & 6670.21$\pm$0.06 & 0.345$\pm$0.001 \\
               & [O{\sc iii}] & 6734.62$\pm$0.02 & 0.345$\pm$0.001 \\
47 & [O{\sc ii}] &  5341.68$\pm$0.06 & 0.433$\pm$0.001 \\
               & H$\beta$    &  6957.91$\pm$0.08 & 0.431$\pm$0.001 \\
               & [O{\sc iii}] & 7098.02$\pm$1.77 & 0.431$\pm$0.001 \\
               & [O{\sc iii}] & 7169.62$\pm$1.19 & 0.432$\pm$0.001 \\
26 & Mg{\sc ii}  &   5546.48$\pm$1.00 & 0.982$\pm$0.001 \\
18 & Mg{\sc ii}   &  5141.63$\pm$0.74 & 0.837$\pm$0.001 \\
\noalign{\smallskip}\hline
\end{tabular}
\end{center}
\label{tab:lines}
\end{table}

\onecolumn

\begin{landscape}
{\tiny
\begin{longtable}{cccccccccc}
\caption{Identified candidates for obscured AGN.}
\label{taobscured}
\\
\hline\hline
Designation & R.A. & Dec. & Error & F$_{x}$ (0.2--12 keV) & eF$_{x}$ & HR2 & HR3 & HR4 & {\it XMM}\_id \\
--          & $\circ$ & $\circ$ & $\arcsec$ &  \ergcm{-11} & erg cm$^{-2}$ s$^{-1}$ & -- & -- & -- & --\\

\hline
\endfirsthead

\multicolumn{10}{c}%
{{\bfseries \tablename\ \thetable{} -- continued from previous page}} \\
\hline
Designation & R.A. & Dec. & Error & F$_{x}$ (0.2--12 keV) & eF$_{x}$ & HR2 & HR3 & HR4 & {\it XMM}\_id \\
--          & $\circ$ & $\circ$ & $\arcsec$ &  \ergcm{-11} & erg cm$^{-2}$ s$^{-1}$ & -- & -- & -- & --\\
\hline
\endhead

\hline \multicolumn{10}{l}{{\tiny Continued on next page}} \\ 
\endfoot

\hline 
\endlastfoot

J002523.92$-$712322.4	&	6.3497	&	$-$71.3896	&	0.1059	&	0.0065	&	0.0010	&	0.055	&	$-$0.42	&	0.27	&	-	\\
J002606.42$-$711301.6	&	6.5267	&	$-$71.2171	&	0.1505	&	0.0042	&	0.0007	&       $-$0.490	&	$-$0.54	&	0.43	&	-	\\
J002636.61$-$712409.5	&	6.6525	&	$-$71.4027	&	0.1365	&	0.0022	&	0.0002	&	0.180	&	$-$0.80	&	$-$0.06	&	-	\\
J004226.41$-$730417.5	&	10.6101	&	$-$73.0716	&	0.0558	&	0.0035	&	0.0005	&       $-$0.200	&	   0.03	&	0.43	&	J004226.3$-$730418	\\
J004633.42$-$734443.2	&	11.6393	&	$-$73.7453	&	0.1022	&	0.0027	&	0.0005	&	0.460	&	$-$0.28	&	0.23	&	J004632.9$-$734444	\\
J005231.59$-$704705.5	&	13.1317	&	$-$70.7849	&	0.1003	&	0.0047	&	0.0010	&	0.360	&	$-$0.69	&	0.34	&	J005232.3$-$704708	\\
J005334.75$-$750404.0	&	13.3948	&	$-$75.0678	&	0.0633	&	0.0100	&	0.0020	&	0.063	&	$-$0.57	&	0.41	&	J005334.5$-$750406	\\
J005415.32$-$743153.0	&	13.5638	&	$-$74.5314	&	0.1259	&	0.0037	&	0.0003	&	0.053	&	$-$0.54	&	0.19	&	J005415.5$-$743153	\\
J005749.70$-$732547.0	&	14.4571	&	$-$73.4297	&	0.0433	&	0.0028	&	0.0005	&	0.095	&	$-$0.23	&	0.55	&	J005750.2$-$732542	\\
J005859.72$-$711457.2	&	14.7489	&	$-$71.2492	&	0.0654	&	0.0079	&	0.0020	&	0.230	&	$-$0.58	&	0.10	&	J005859.7$-$711458	\\
J010033.38$-$715037.2	&	15.1391	&	$-$71.8437	&	0.1435	&	0.0024	&	0.0005	&       $-$0.009	&	$-$0.43	&	0.45	&	J010033.4$-$715039	\\
211.16703.311	        &	15.5600	&	$-$73.2739	&	1	&	0.0054	&	0.0010	&	0.480	&	$-$0.84	&	0.71	&	-	\\
211.16765.212	        &	15.6450	&	$-$72.9061	&	1	&	0.0045	&	0.0010	&	0.690	&	$-$0.58	&	0.12	&	J010234.7$-$725425	\\
J011420.98$-$722328.4	&	18.5874	&	$-$72.3912	&	0.1812	&	0.0030	&	0.0004	&	0.450	&	$-$0.55	&	0.22	&	J011421.3$-$722327	\\
J011606.88$-$731725.0	&	19.0287	&	$-$73.2903	&	0.0981	&	0.0038	&	0.0009	&	1	&	$-$0.36	&	0.46	&	J011606.7$-$731725	\\
J011732.79$-$735228.6	&	19.3867	&	$-$73.8746	&	0.1794	&	0.0042	&	0.0020	&	0.340	&	$-$0.38	&	0.34	&	-	\\
J012024.93$-$734720.3	&	20.1039	&	$-$73.7890	&	0.1024	&	0.0030	&	0.0007	&	0.430	&	$-$0.38	&	0.15	&	-	\\
J012938.05$-$742144.8	&	22.4086	&	$-$74.3625	&	0.1208	&	0.0070	&	0.0010	&	0.040	&	$-$0.61	&	0.27	&	-	\\
J013119.92$-$742505.7	&	22.8330	&	$-$74.4183	&	0.1324	&	0.0028	&	0.0005	&	0.190	&	$-$0.84	&	0.70	&	-	\\
J013254.61$-$743954.3	&	23.2276	&	$-$74.6651	&	0.0870	&	0.0071	&	0.0020	&	0.500	&	$-$0.49	&	0.40	&	-	\\

\hline\hline
\end{longtable}
}

 \begin{center}

  { \scriptsize The designation, RA, Dec, associated error and the X and Y colours correspond to the ALLWISE or HMQ/MILLIQUAS source.}\\
  
  {\scriptsize The X-ray flux and the associated error (F$_{x}$ and eF$_{x}$) are from the corresponding X-ray counterpart.}\\
  
   {\scriptsize HR2 HR3 and HR4 refer to the hardness ratios in the energy range of 0.5--1, 1--2 and 2--4.5 keV
respectively.}\\

  {\scriptsize {\it XMM}\_ids correspond to sources already included in \cite{2013A&A...558A...3S}.}

\end{landscape}

\begin{landscape}
{\tiny
\begin{longtable}{ccccccccccc}
\caption{Sample of the first five entries in the catalogue of AGN compiled from this work.}
\label{tacat}
\\
\hline\hline
Designation & R.A. & Dec. & Error & R.A. & Dec. & Error & separation & X & Y & F$_{x}$ (0.2--12 keV) \\
 -- & $\circ$ & $\circ$ & $\arcsec$ & $\circ$ (X-ray) & $\circ$ (X-ray) & $\arcsec$ (X-ray) & $\arcsec$ & $\log \frac{(f12\mu m)}{(f4.6 \mu m)}$ & $\log \frac{(f4.6\mu m)}{(f3.4 \mu m)}$ & \ergcm{-11} \\
 \hline
 \endfirsthead
\multicolumn{11}{c}
{{\bfseries \tablename\ \thetable{} -- continued from previous page}}\\
 \hline
 Designation & R.A. & Dec. & Error & R.A. & Dec. & Error & separation & X & Y & F$_{x}$ (0.2--12 keV) \\
  & $\circ$ & $\circ$ & $\arcsec$ & $\circ$ (X-ray) & $\circ$ (X-ray) & $\arcsec$ (X-ray) & $\arcsec$ & $\log \frac{(f12\mu m)}{(f4.6 \mu m)}$ & $\log \frac{(f4.6\mu m)}{(f3.4 \mu m)}$ & \ergcm{-11} \\
 \hline
 \endhead
 \hline \multicolumn{11}{l}{{\tiny Continued on next page}} \\ 
 \endfoot
 \hline 
 \endlastfoot
 J002330.72$-$722043.6 & 5.87803	& $-$72.3455	&	0.0374	&	5.87776	&	$-$72.3455	&	0.341	&	0.29	&	0.454	&	0.046	&	0.2000	\\
 J002516.40$-$712456.4 & 6.31835	& $-$71.4157	&	0.0986	&	6.31832	&	$-$71.4156	&	0.879	&	0.15	&	0.441	&	0.232	&	0.0088	\\ 
 J002523.92$-$712322.4 & 6.3497	& $-$71.3896	&	0.1059	&	6.34863	&	$-$71.3895	&	1.240	&	1.30	&	0.244	&	0.152	&	0.0065	\\
 J002557.21$-$711617.2 & 6.4884	& $-$71.2715	&	0.0719	&	6.48785	&	$-$71.2713	&	0.694	&	0.96	&	0.460	&	0.303	&	0.0064	\\
 J002557.82$-$712540.7 & 6.49095	& $-$71.4280	&	0.0924	&	6.49109	&	$-$71.4283	&	1.160	&	1.20	&	0.391	&	0.076	&	0.0026 \\
 \end{longtable}
\begin{longtable}{ccccccccccc}
%
\hline\hline
Designation & eF$_{x}$ & HR1 & HR2 & HR3 & HR4 & $Y-J$ & $J-K_{\rm s}$ & {\it XMM}\_id & $z$ & Identification \\
 -- & erg cm$^{-2}$ s$^{-1}$ & -- & -- & -- & -- & mag & mag & --& -- & -- \\
\hline
\endfirsthead
\multicolumn{11}{c}
{series \tablename\ \thetable{} -- continued from previous page} \\
 \hline
Designation & eF$_{x}$ & HR1 & HR2 & HR3 & HR4 & $Y-J$ & $J-K_{\rm s}$ & {\it XMM}\_id & $z$ & Identification \\
 -- & erg cm$^{-2}$ s$^{-1}$ & -- & -- & -- & -- & mag & mag & --& -- & -- \\
\hline
\endhead
\hline \multicolumn{11}{l}{{\tiny Continued on next page}} \\ 
\endfoot
\hline 
\endlastfoot
J002330.72$-$722043.6   & 0.0020	&	0.296	&       $-$0.080	&	$-$0.458	&	$-$0.504	&	0.617	&	1.35	&	-	&	-	&	ALLWISE	\\
J002516.40$-$712456.4	& 0.0010	&	0.114	&	0.069	&	$-$0.155	&	$-$0.216	&	0.415	&	1.32	&	-	&	-	&	ALLWISE	\\
J002523.92$-$712322.4	& 0.0010	&	0.330	&	0.055	&	$-$0.425	&	   0.271	&	0.619	&	1.29	&	-	&	-	&	-	\\
J002557.21$-$711617.2	& 0.0005	&	0.187	&       $-$0.080	&	$-$0.577	&	$-$0.699	&	0.098	&	1.14	&	-	&	-	&	-	\\
J002557.82$-$712540.7	& 0.0004	&       $-$0.018	&    $-$0.019	&	$-$0.339	&	$-$0.183	&	0.748	&	1.32	&	-	&	-	&	HMQ/MILLIQUAS	\\
\label{tacat}
\end{longtable}
\begin{center}
{ \scriptsize Columns have the same meanings as in Tables~\ref{tab:new} and \ref{taobscured}. $z$ refers to the redshift if known.}\\
\end{center}
%
%
 }
\end{landscape}
\end{landscape}

\section{Figures}


\begin{figure*}
\centering
\subfigure[Candidate 59]{\includegraphics[width=60mm]{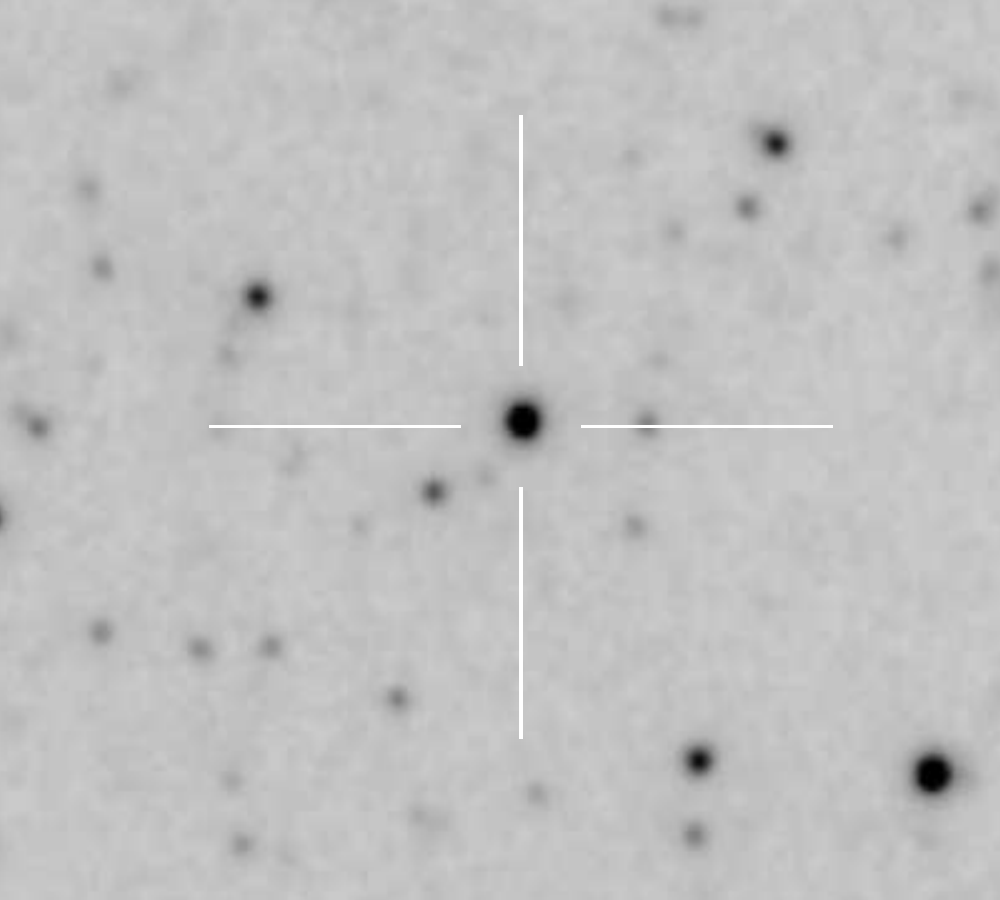}} 
\subfigure[Candidate 35]{\includegraphics[width=60mm]{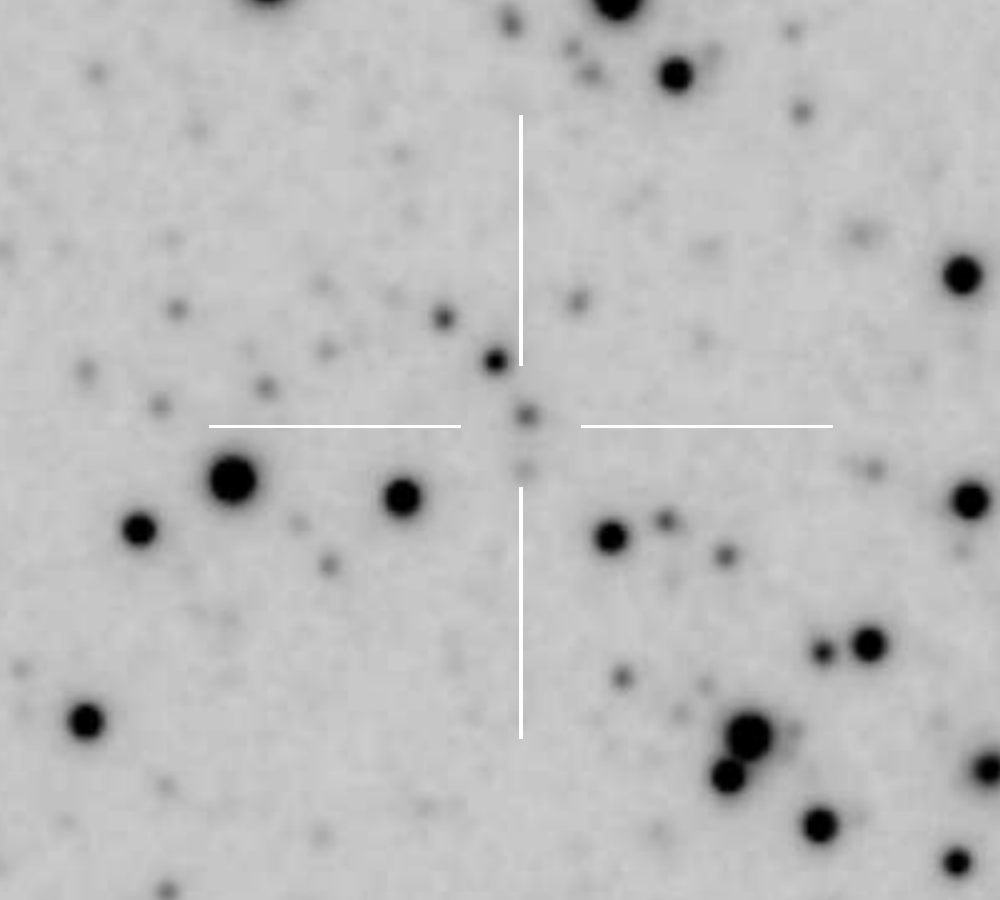}}
\subfigure[Candidate 06]{\includegraphics[width=60mm]{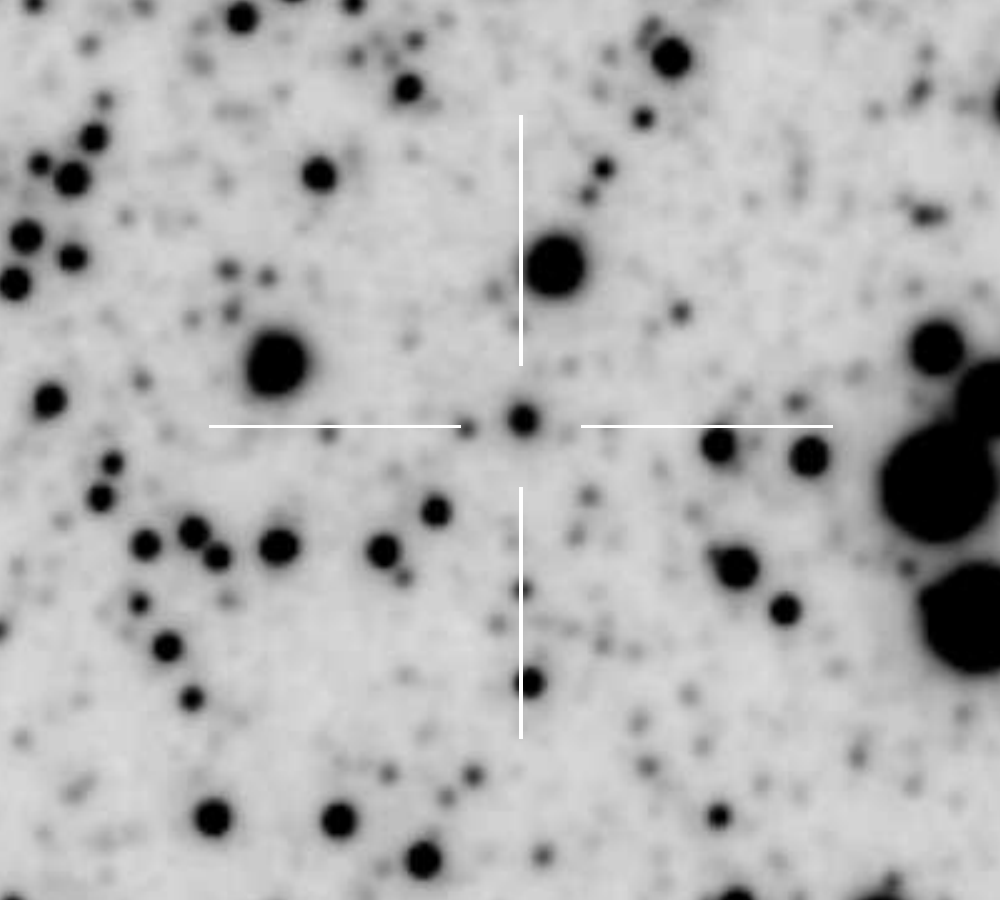}}
\subfigure[Candidate 47]{\includegraphics[width=60mm]{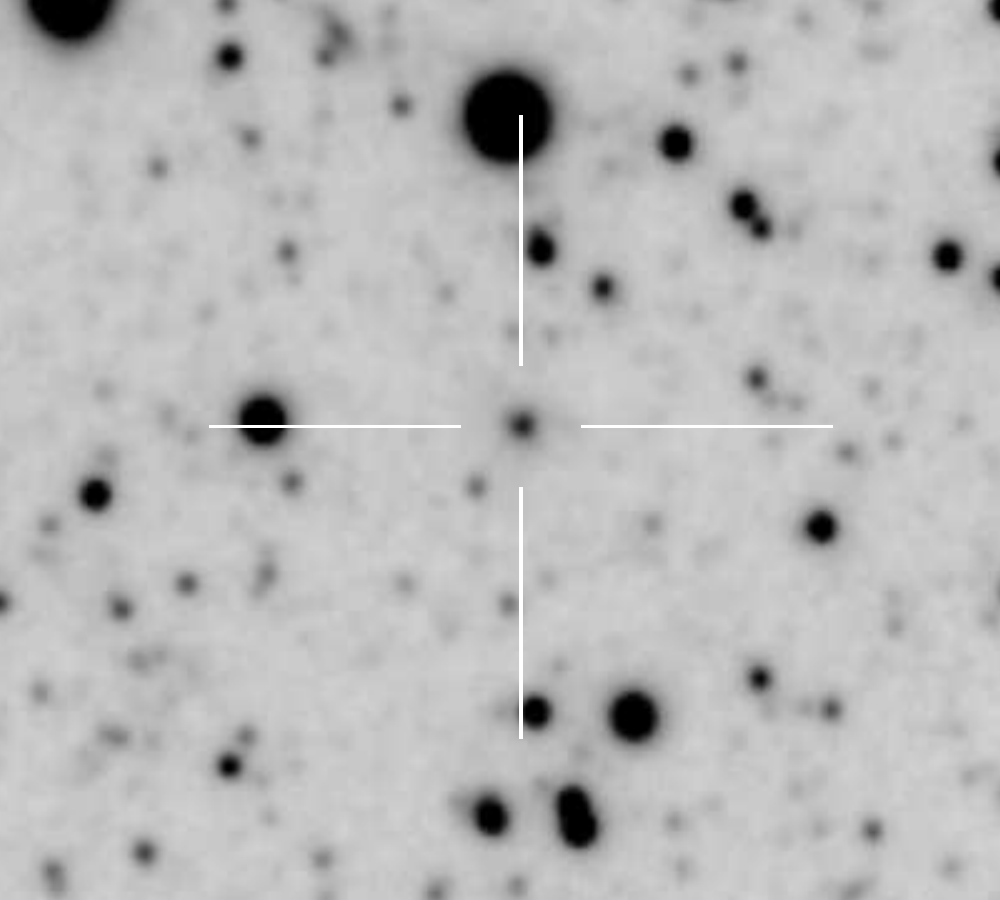}}
\subfigure[Candidate 26]{\includegraphics[width=60mm]{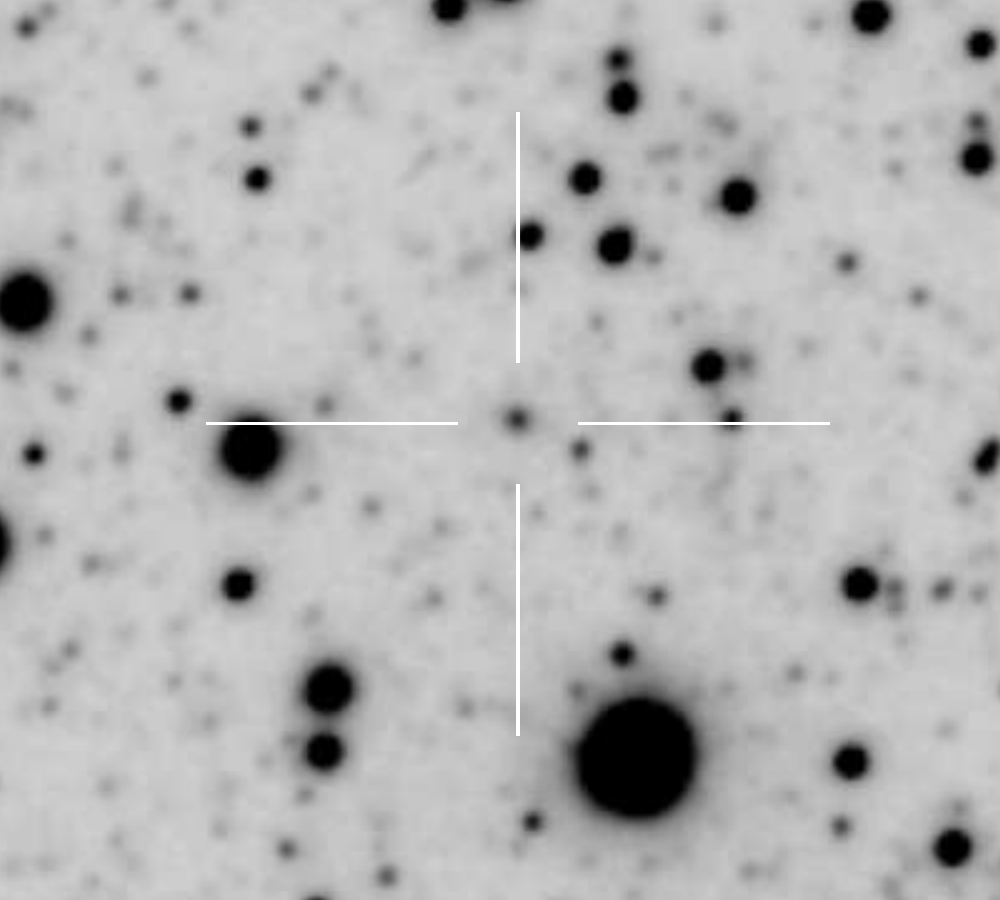}}
\subfigure[Candidate 18]{\includegraphics[width=60mm]{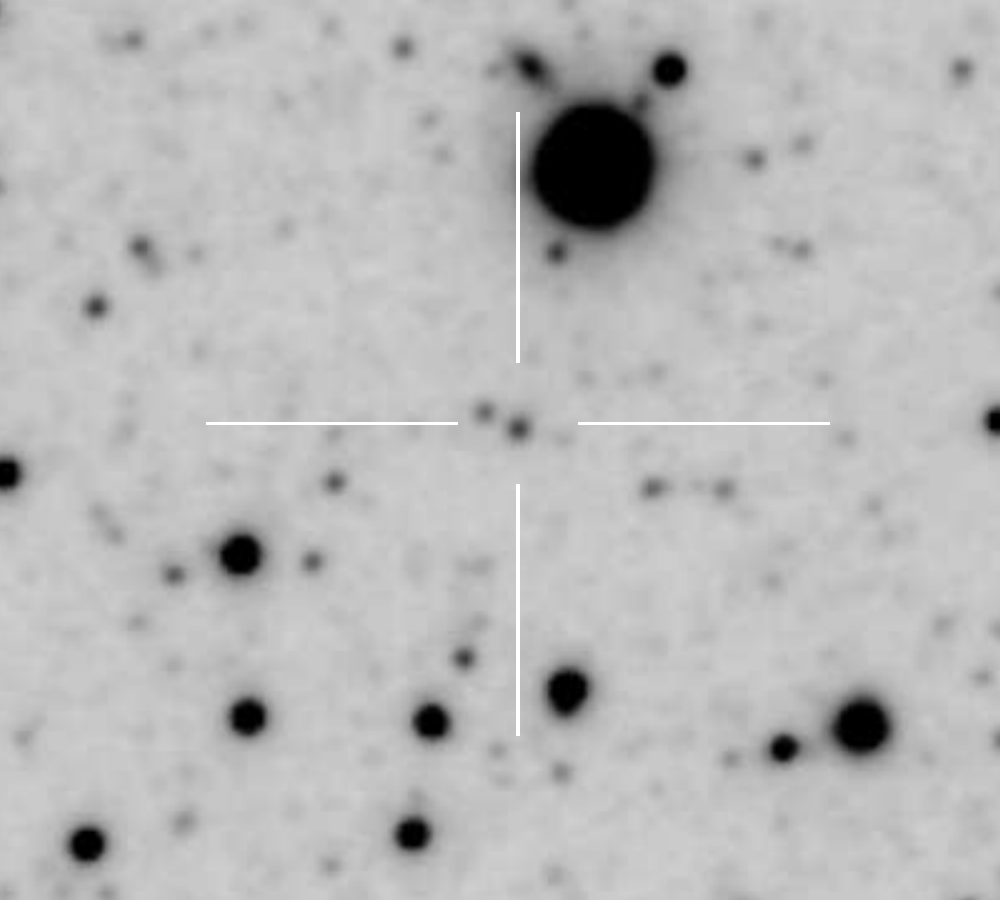}}
\caption{Finding charts in the $Y$-band for all 6 objects (crosses) 
with follow-up spectroscopy. The images are 1$\times$1\,arcmin$^2$. 
North is at the top and east is to the left.}
\label{fig:fcs}
\end{figure*}

\end{appendix}

\Online

\end{document}